\documentclass[twocolumn,showpacs,preprintnumbers,amsmath,amssymb]{revtex4}

\usepackage{graphicx}
\usepackage{dcolumn}
\usepackage{bm}

\def\bea{\begin{eqnarray}}
\def\eea{\end{eqnarray}}

\begin{document} 

\preprint{Version 1.6}

\title{The azimuth structure of nuclear collisions -- I}

\author{Thomas A. Trainor and David T. Kettler}
\address{CENPA 354290, University of Washington, Seattle, WA  98195}


\date{\today}

\begin{abstract}
We describe azimuth structure commonly associated with elliptic and directed flow in the context of 2D angular autocorrelations for the purpose of precise separation of so-called nonflow (mainly minijets) from flow. We extend the Fourier-transform description of azimuth structure to include power spectra and autocorrelations related by the Wiener-Khintchine theorem. We analyze several examples of conventional flow analysis in that context and question the relevance of reaction plane estimation to flow analysis. We introduce the 2D angular autocorrelation with examples from data analysis and describe a simulation exercise which demonstrates precise separation of flow and nonflow using the 2D autocorrelation method. We show that an alternative correlation measure based on Pearson's normalized covariance provides a more intuitive measure of azimuth structure.
\end{abstract}


\maketitle

 \section{Introduction}


A major goal of the RHIC is production of color-deconfined or QCD matter in heavy ion (HI) collisions, a bulk QCD medium extending over a nontrivial space-time volume which is in some sense thermalized and whose dynamics are dominated in some sense by quarks and gluons as the dominant degrees of freedom~\cite{theoryequil}. ``Matter'' in this context means an aggregate of constituents  in an equilibrium state, at least locally in space-time, such that thermodynamic state variables provide a nearly complete description of the system. Demonstration of thermalization is seen by many as a necessary part of the observation of QCD matter. 
 
\subsection{Global variables}

One method proposed to demonstrate the existence of QCD matter is to measure trends of {\em global event variables}, statistical measures formulated by analogy with macroscopic thermodynamic quantities and based on integrals of particle yields over kinematically accessible momentum space. E.g., temperature analogs include spectrum inverse slope parameter $T$ and ensemble-mean $p_t$ $\hat p_t$. Chemical analogs include particle yields and their ratios, such as the ensemble-mean $K/\pi$ ratio~\cite{global}. Corresponding fluctuation measures have been formulated for the event-wise mean $p_t$ $\langle p_t \rangle$ (``temperature'' fluctuations) and  $K/\pi$ ratio (chemical or flavor fluctuations)~\cite{stock,phipt,cltpaper}. Arguments by analogy are less appropriate when dealing with small systems (`small' in particle number, space and/or time) where large deviations from macroscopic thermodynamics may be encountered.
 
\subsection{Flow analysis}

One such global feature is the large-scale angular structure of the event-wise particle distribution. The components of angular structure described by low-order spherical or cylindrical harmonics are conventionally described as ``flows.'' The basic assumption is that such structure represents {\em collective} motion of a thermalized medium, and hydrodynamics is therefore an appropriate description. Observation of larger flow amplitudes is therefore interpreted by many to provide direct evidence for event-wise thermalization in heavy ion collisions~\cite{flowequil}. Given those assumptions each collision event is treated separately. Event-wise angular distributions are fitted with model functions associated with collective dynamics. The model parameters are interpreted physically in a thermodynamic (i.e., collective, thermalized) context. 

However, collective flow in many-body physics is a complex topic with longstanding open issues. There is conflict in the description of nuclear collisions between continuum hydrodynamics and discrete multiparticle systems which echoes the state of physics prior to the study of Brownian motion by Einstein and Perrin~\cite{brownian,perrin}. Beyond the classical dichotomy between discrete and continuous dynamics there is the still-uncertain contribution of quantum mechanics to the early stages of nuclear collisions. Quantum transitions may play a major role in phenomena perceived to be ``collective.'' Premature imposition of hydrodynamic (hydro) models on collision data may hinder full understanding.

\subsection{Nonflow and multiplicity distortions}


A major concern for conventional flow analysis is the presence of ``nonflow,'' non-sinusoidal contributions to azimuth structure often comparable in amplitude to sinusoid amplitudes (flows). Nonflow is treated as a systematic error in flow analysis, reduced to varying degrees by analysis strategies. Another significant systematic issue is ``multiplicity distortions'' associated with small event multiplicities, also minimized to some degree by analysis strategies. Despite corrections nonflow and small multiplicities remain a major limitation to conventional flow measurements. For those reasons flow measurements in peripheral heavy ion collisions are typically omitted. The opportunity is then lost to connect the pQCD physics of elementary collisions to nonperturbative, possibly collective dynamics in heavy ion collisions.

\subsection{Minijets}

A series of recent experiments has demonstrated that the nonsinusoidal components of angular correlations at full RHIC energy are dominated by fragments from low-$Q^2$ partons or {\em minijets}~\cite{ppcorr,axialci,ptscale,edep}. Minijets in RHIC p-p and A-A collisions have been studied extensively {\em via} fluctuations~\cite{ptscale,edep} and two-particle correlations~\cite{ppcorr,axialci}. Minijets may dominate the production mechanism for the QCD medium~\cite{mueller}, and may also provide the best probe of medium properties, including the extent of thermalization. Comparison of minijets in elementary and heavy ion collisions may relate medium properties and collective motion to a theoretical QCD context.

Ironically, demonstrating the existence and properties of collective flows and of jets (collective hadron motion from parton collisions and fragmentation) is formally equivalent. Identifying a partonic ``reaction plane'' and determining a nucleus-nucleus reaction plane require similar techniques. For example, sphericity has been used to obtain evidence for deformation of particle/$p_t$/$E_t$ angular distributions due to parton collisions~\cite{spherpart} and collective nucleon flow~\cite{spherflow}. At RHIC we should ask whether final-state angular correlations depend on the geometry of parton collisions (minijets), N-N collisions or nucleus-nucleus collisions (flows), or all three. The analogy is important because to sustain a flow interpretation one has to prove that there is a difference: e.g., to what extent do parton collisions contribute to flow correlations or mimic them? The phenomena coexist on a continuum. 

To resolve such ambiguities we require analysis methods which treat flow and minijets on an equal footing and facilitate their comparison, methods which do not impose the hypothesis to be tested on the measurement scheme. To that end we should: 1)  develop a consistent set of neutral symbols; 2) manipulate random variables with minimal approximations; 3) introduce proper statistical references so that nonstatistical correlations of any origin can be isolated  unambiguously; 4) treat azimuth structure {\em ab initio} in a model-independent manner using standard mathematical methods (e.g., standard Fourier analysis); and 5) include what is known about minijets (``nonflow'') and ``flow'' in a more general analysis based on two-particle correlations. 
 
\subsection{Structure of this paper}

An underlying theme of this paper is the formal relation between event-wise azimuth structure in nuclear collisions and Brownian motion, and how that relation can inform our study of heavy ion collisions. We begin with a review of Fourier transform theory and the relation between power spectra and autocorrelations. That material forms a basis for analysis of sinusoidal components of angular correlations in nuclear collisions which is well-established in standard mathematics. 

We then review the conventional methods of flow analysis from Bevalac to RHIC. Five papers are discussed in the context of Fourier transforms, power spectra and autocorrelations. To facilitate a more general description of angular asymmetries we set aside flow terminology (except as required to make connections with the existing literature) and move to a model-independent description in terms of {\em spherical and cylindrical multipole moments}. We emphasize the relation of ``flows'' to multipole moments as model-independent correlation measures. Physical interpretation of multipole moments  is an open question.

We then consider whether event-wise estimation of the reaction plane is necessary for ``flow'' studies. The conventional method of flow analysis is based on such estimation, assuming that event-wise statistics are required to demonstrate collectivity, and hence thermalization, in heavy ion collisions.

We define the 2D joint angular autocorrelation and describe its properties. The autocorrelation is fundamental to time-series analysis, the Brownian motion problem and its generalizations and astrophysics, among many other fields. It is shown to be a powerful tool for separating ``flow'' from ``nonflow.'' The autocorrelation eliminates biases in conventional flow analysis stemming from finite multiplicities, and makes possible bias-free study of centrality variations in A-A collisions down to N-N collisions.

``Nonflow'' is dominated by minijets (minimum-bias parton fragments, mainly from low-$Q^2$ partons) which can be regarded as Brownian probe particles for the QCD medium, offering the possibility to explore small-scale medium properties. Minijet systematics provide strong constraints on ``nonflow'' in the conventional flow context. Interaction of minijets with the medium, and particularly its collective motion, is the subject of paper II of this series.

Finally, we consider examples from RHIC data of autocorrelation structure. We show the relation between ``flow'' and minijets, how conventional flow analysis is biased by the presence of minijets, and how the autocorrelation method eliminates that bias and insures accurate separation of different collision dynamics.

We include several appendices. In App. A we review Brownian motion and its formal connection to azimuth correlations in nuclear collisions. In App. B we review the algebra of random variables in relation to conventional flow analysis techniques. We make no approximations in statistical analysis and invoke proper correlation references to obtain a minimally-biased, self-consistent analysis system in which flow and nonflow are precisely distinguished. In App. C we review the mathematics of spherical and cylindrical multipoles and sphericity. In App. D we review subevents, scalar products and event-plane resolution. In App. E we summarize some A-A centrality issues related to azimuth multipoles and minijets.

\section{Fourier Analysis} \label{foorier}

The azimuth structure of nuclear collisions is part of a larger problem: angular correlations of number, $p_t$ and $E_t$ on angular subspace $(\eta_1,\eta_2,\phi_1,\phi_2)$. There is a formal similarity between event-wise particle distributions on angle and the time series of displacements of a particle in  Brownian motion. In either case the distribution is discrete, combining a large random component with the possibility of a smaller deterministic component. The mathematical description of Brownian motion includes as a key element the autocorrelation density, related to the Fourier power spectrum through the Wiener-Khintchine theorem ({\em cf.} App.~\ref{brown}). 

The Fourier series describes arbitrary distributions on bounded angular interval $2\pi$ or distributions periodic on an unbounded interval. The azimuth particle distribution from a nuclear collision is drawn from (samples) a combination of sinusoids nearly invariant on rapidity near midrapidity, conventionally described as ``flows,'' and other azimuth structure localized on rapidity and conventionally described as ``nonflow.'' The two contributions are typically comparable in amplitude.

We first consider the mathematics of the Fourier transform and power spectrum and their role in conventional flow analysis~\cite{fourtrans}. We assume for simplicity that the only angular structure in the data is represented by a few lowest-order Fourier terms. In conventional flow analysis the azimuth distribution is described solely by a Fourier series, and corrections are applied in an attempt to compensate for ``nonflow'' as a systematic error. We later return to the more general angular correlation problem and consider non-sinusoidal (nonflow) structure described by non-Fourier model functions in the larger context of  2D (joint) angular autocorrelations. Precise description of the composite structure requires a hybrid mathematical model.

Event-wise random variables are denoted by a tilde. Variables without tildes are ensemble averages, indicated in some cases explicitly by overlines. Event-wise averages are indicated by angle brackets. The algebra of random variables is discussed in App.~\ref{stats}. Where possible we employ notation consistent with conventional flow analysis.

\subsection{Azimuth densities}
 
The event-wise azimuth density (particle, $p_t$ or $E_t$) is a set of $n$ samples from a parent density integrated over some $(p_t,\eta)$ acceptance, a sum over Dirac delta functions (particle positions)
\bea
\tilde \rho(\phi) = \sum_{i=1}^n r_i\, \delta(\phi - \phi_i)
\eea 
The $r_i$ are weights (1, $p_t$ or $E_t$) appropriate to  a given physical context. We assume integration over one unit of pseudorapidity, so multiplicity $n$ estimates $dn/d\eta$ (similarly for $p_t$ and $E_t$). The continuum parent density, not directly observable, is the object of analysis.  Fixed parts of the parent density are estimated by a histogram averaged over an event ensemble. The discrete nature of the sample distribution and its statistical character present analysis challenges which are one theme of this paper. 

The correlation structure of the single-particle azimuth density is manifested in ensemble-averaged multiparticle (two-particle, etc.) densities. Accessing that {structure} by projection of multiparticle spaces to 2D or 1D with minimal distortion is the object of {\em correlation analysis}. In each event the two-particle density is the Cartesian product $\tilde \rho(\phi_1,\phi_2) = \tilde \rho(\phi_1)\, \tilde \rho(\phi_2)$ 
 \bea
 \tilde \rho(\phi_1,\phi_2) &=& \sum_{i=1}^n r_i^2 \, \delta(\phi_1 - \phi_i) \delta(\phi_2 - \phi_i) \\ \nonumber
&+& \sum_{i\neq j}^{n,n-1} r_i r_j\, \delta(\phi_1 - \phi_i) \delta(\phi_2 - \phi_j),
 \eea
where the first term represents {\em self pairs}. Ensemble-averaged two-particle distribution $\rho(\phi_1,\phi_2)$ with correlations is not generally factorizable. By comparing the averaged two-particle distribution to a factorized (or mixed-pair) statistical reference two-particle correlations are revealed. In a later section we compare multipole moments from two-particle correlation analysis on azimuth to results from conventional flow analysis methods.

\subsection{Fourier transforms on azimuth}

A Fourier series is an efficient representation if azimuth structure approximates a constant plus a few sinusoids whose wavelengths are integral fractions of $2\pi$. A Fourier representation of a peaked distribution (e.g., jet cone) is not an efficient representation.  We assume a simple combination of the lowest few Fourier terms.

The Fourier forward transform (FT) is
\bea \label{fourier}
\tilde \rho(\phi) 
&=&   \sum_{m = -\infty}^{\infty} \frac{\tilde {\bf Q}_m}{2\pi} \exp(i\, m \phi) \\ \nonumber
 &=&   \frac{\tilde Q_0}{2\pi} +   2\sum_{m = 1}^{\infty} \frac{\tilde Q_m}{2\pi} \cos( m [\phi - \Psi_m]),
\eea  
where boldface $\tilde {\bf Q}_m$ is an event-wise complex amplitude, $\tilde Q_m$ is its magnitude and $\Psi_m$ its phase angle. The second line arises because $\tilde \rho(\phi)$ is a real function, and the practical upper limit on $m$ is particle number $n$ (wavelength $\sim$ mean interparticle spacing).  $\tilde Q_m / 2\pi = \tilde\rho_m$ is the amplitude of the density variation associated with the $m^{th}$ sinusoid. With $r_i \rightarrow 1$ $\tilde Q_m$ is the corresponding number of particles in $2\pi$ if that density were uniform.  $\Psi_m$ is event-wise by definition and does not require a tilde. The reverse transform (RT) is
\bea \label{freverse}
\tilde {\bf Q}_m &=&   \int_{-\pi}^{\pi}\hspace{-.1in} d\phi \, \tilde \rho(\phi ) \exp(-i\, m \phi ) \\ \nonumber
&=&  \sum_i^n r_i \exp(-i\, m \phi_i) \\ \nonumber
&=& \tilde Q_m\, \exp(i\, m\Psi_m).
\eea

For the discrete transform $\phi \in [-\pi,\pi]$ is partitioned into $M$ equal bins with bin contents  $\tilde r_l$ and bin centers $\phi_l$. We multiply Eq.~(\ref{fourier}) by bin width $\delta \phi = 2\pi / M$, and the Fourier transform pair becomes
\bea
 \tilde r_l  &=&   \frac{\tilde Q_0}{M} +   2\sum_{m = 1}^{M/2} \frac{\tilde Q_m}{M} \cos( m [\phi_l - \Psi_m]) \\ \nonumber
{\tilde {\bf Q}_m}  &=& \sum_{l = -M/2}^{M/2}  \tilde r_l\, \exp(-i\,m \phi_l),
\eea
where the $\tilde r_l$ are also random variables. The upper limit $M/2$ in the first line is a manifestation of the {\em Nyquist sampling theorem}~\cite{nyquist}. With $r_i \rightarrow 1$ $\tilde r_l \rightarrow \tilde n_l$ and $Q_m / M$ is the maximum number of particles in a bin associated with the $m^{th}$ sinusoid.

\subsection{Autocorrelations and power spectra}

The azimuth autocorrelation density $\rho_A(\phi_\Delta)$ is a {\em projection by averaging} of the pair density on two-particle azimuth space $(\phi_1,\phi_2)$ onto difference axis $\phi_\Delta = \phi_1 - \phi_2$. The autocorrelation concept is not restricted to periodic or bounded distributions or discrete Fourier transforms~\cite{fourtrans}. In what follows $\tilde \rho_A$ includes self pairs. 
The autocorrelation density is defined as~\cite{inverse}
\bea
\tilde \rho_A(\phi_\Delta) &\equiv& \frac{1}{2\pi} \int_{-\pi}^\pi d\phi\, \tilde \rho(\phi)\, \tilde \rho(\phi + \phi_\Delta) \\ \nonumber
 &=& \frac{1}{2\pi} \sum_{i,j=1}^{n} r_i r_j \int_{-\pi}^\pi d\phi\,  \delta(\phi - \phi_i)\,  \delta(\phi - \phi_j + \phi_\Delta) \\ \nonumber
 &=& \frac{1}{2\pi} \sum_{i,j=1}^{n}  r_i r_j \,   \delta(\phi_i - \phi_j + \phi_\Delta). \\ \nonumber
\eea
Using Eq.~(\ref{fourier}) we obtain the FT as
\bea
\tilde \rho_A(\phi_\Delta)&=&  \frac{1}{2\pi} \int_{-\pi}^\pi d\phi\, \times \\ \nonumber
 &&\sum_{m=-\infty}^{\infty} \frac{\tilde {\bf Q}_m}{2\pi} \exp(i\, m \phi) \times \\ \nonumber
 &&\sum_{m'=-\infty}^{\infty} \frac{\tilde {\bf Q}^*_{m'}}{2\pi} \exp(-i\, m' [\phi + \phi_\Delta])  \\ \nonumber
&=&   \sum_{m=-\infty}^{\infty} \frac{\tilde Q^2_m}{ [2\pi]^2} \exp(-i\ m \phi_\Delta) \nonumber  \\ \nonumber
&=&  \frac{\tilde Q_0^2}{ [2\pi]^2} + 2\sum_{m=1}^{\infty} \frac{\tilde Q^2_m}{ [2\pi]^2} \cos( m \phi_\Delta),
\eea
and the RT as
\bea  \label{powspec}
{\tilde Q^2_m}  &=& n^2 \langle r \cos(m [\phi - \Psi_m])\rangle^2  \\ \nonumber
&=& {2\pi}   \int_{-\pi}^\pi d\phi_\Delta \, \tilde \rho_A(\phi_\Delta)\cos( m \phi_\Delta) \\ \nonumber
&=& \sum_{i,j=1}^{n} r_i r_j \,    \cos(m[\phi_i - \phi_j] ) \\ \nonumber
&=& n\langle r^2 \rangle + n(n-1)\langle r^2  \cos(m \phi_\Delta) \rangle. \\ \nonumber
&=& n \langle r^2\rangle + n(n-1) \langle r^2 \cos^2(m[\phi - \Psi_r])\rangle.
\eea
Phase angle $\Psi_m$ has been eliminated, and $\Psi_r$ will be identified with the {\em reaction plane} angle. 

We can write the same relations for ensemble-averaged quantities because the terms are positive-definite
\bea \label{fortran}
\rho_A(\phi_\Delta) &\equiv& \frac{ Q_0^2}{ [2\pi]^2} + 2\sum_{m=1}^{\infty} \frac{ Q^2_m}{ [2\pi]^2} \cos( m \phi_\Delta),
\eea
with RT
\bea \label{wienkh}
{ Q^2_m} &=& {2\pi}   \int_{-\pi}^\pi d\phi_\Delta \,  \rho_A(\phi_\Delta)\cos( m \phi_\Delta) \\ \nonumber
&=& \overline{\sum_{i,j=1}^{n}  r_i r_j \,    \cos(m[\phi_i - \phi_j] )} \\ \nonumber
&=& \overline{n \langle r^2\rangle} + \overline{n(n-1)\langle r^2 \cos(m \phi_\Delta) \rangle}.
 \eea
We have adopted the convention $Q^2_m = \overline{\tilde Q^2_m}$ to lighten the notation. Coefficients $Q_m^2$ are  {\em power-spectrum} elements on wave-number index $m$. That FT transform pair expresses the {\em Wiener-Khintchine theorem} which relates power-spectrum elements $Q^2_m$ to  autocorrelation density $\rho_A(\phi_\Delta)$. The autocorrelation provides precise access to two-particle correlations given enough collision events, {\em no matter how small the event-wise multiplicities}. 

\subsection{Autocorrelation structure}

The autocorrelation concept was developed in response to the Brownian motion problem and the Langevin equation, a differential equation describing Brownian motion which contains a stochastic term. The concept is already apparent in Einstein's first paper on the subject ({\em cf.} App.~\ref{brown}). The large-scale, possibly-deterministic motion of the Brownian probe particle must be separated from its small-scale random motion due to thermal collisions with molecules. Similarly, we  want to extract azimuth correlation structure persisting in some sense over an event ensemble from event-wise random variations. The autocorrelation technique is designed for that purpose.

The statistical reference for a power spectrum is the {\em white-noise background} representing an uncorrelated system. The reference is typically uniform up to large wave number or frequency (hence {\em white} noise), an inevitable part of any power spectrum from a discrete process (point distribution). The ``signal'' is typically limited to a bounded region (signal bandwidth) of the spectrum at smaller frequencies or wave numbers. 

From Eq.~(\ref{powspec}) the power-spectrum elements are
\bea \label{powspec2}
 \tilde Q_m^2 &=&  \sum_i^n r^2_i + {\sum_{i \neq j}^{n,n-1} r_i r_j \,  \cos( m[\phi_i - \phi_j]) } \\ \nonumber
&=&  n \langle r^2 \rangle + {n(n-1)  \langle r^2 \cos(m\phi_\Delta) \rangle}
\eea
The first term in Eq.~(\ref{powspec2}) is $\tilde Q^2_{ref}$, the white-noise background component of the power spectrum common to all spectrum elements. The second term, which we denote $\tilde V_m^2$,  represents true two-particle  azimuth correlations. Note that $\tilde Q_0^2 = n^2 \langle r^2 \rangle$, whereas $\tilde V_0^2 = n(n-1) \langle r^2 \rangle$.
In terms of complex (or vector) amplitudes we can write
\bea
\tilde {\bf Q}_m 
&=& \tilde  {\bf Q}_{ref} + \tilde {\bf V}_m,
\eea
where $\tilde {\bf Q}_{ref}$ represents a random walker. There is no cross term in Eq.~(\ref{powspec2}) because ${\bf Q}_{ref}$ and ${\bf V}_m$ are uncorrelated.

Inserting the power-spectrum elements into Eq.~(\ref{fortran}) we obtain the {\em ensemble-averaged} autocorrelation density
\bea \label{autocorr}
\rho_A(\phi_\Delta) &=& \frac{\overline{ n\langle r^2\rangle} }{2\pi}\delta(\phi_\Delta) +  \frac{\overline{n(n-1)\langle r^2 \rangle}}{[2\pi]^2}  \\ \nonumber
&+& 2 \sum_{m=1}^{\infty} \frac{V_m^2}{[2\pi]^2} \, \cos(m \phi_\Delta).
 \eea
The first term is the self-pair or statistical noise term, which can be excluded from $\rho_A$ by definition simply by excluding self pairs. The second term, with $V_0^2 = \overline{n(n-1)\langle r^2 \rangle}$, is a uniform component, and the third term is the sinusoidal correlation structure. The self-pair term is referred to in conventional flow analysis as the ``autocorrelation,'' in the sense of a bias or systematic error, but that is a notional misuse of standard mathematical terminology. The true autocorrelation density is the entirety of Eq.~(\ref{autocorr}), including (in  this simplified case) the self-pair term,  the uniform component and the sinusoidal two-particle correlations. 

In general, the single-particle ensemble-averaged distribution $\rho_0$ may be structured on $(\eta,\phi)$. We want to subtract the corresponding reference structure from the two-particle distribution to isolate the true correlations. In what follows we assume $r_i = 1$ for simplicity, therefore describing {\em number} correlations. We subtract factorized reference autocorrelation  $\rho_{A,ref}(\phi_1,\phi_2)  = \rho_0(\phi_1)\, \rho_0(\phi_2)$ representing a system with no correlations, with $\rho_0  =  \bar n / 2\pi\simeq \overline{d^2 n/d\eta d\phi}$ in this simple example, to obtain the difference autocorrelation
\bea
\Delta \rho_A(\phi_\Delta) &=& \rho_A -  \rho_{A,ref} \\ \nonumber
&=& \frac{\sigma^2_n - \bar n}{[2\pi]^2} + 2 \sum_{m=1}^{\infty} \frac{V_m^2}{[2\pi]^2} \, \cos(m\, \phi_\Delta) .
\eea
The first term measures excess (non-Poisson) multiplicity fluctuations in the full $(p_t,\eta,\phi)$ acceptance. The second term is a sum over cylindrical multipoles. We now divide the autocorrelation difference by $ \sqrt{\rho_{A,ref}} = \rho_0 = \bar n / 2\pi$ to form the density ratio
\bea
\frac{\Delta \rho_A}{\sqrt{\rho_{A,ref}}} &=& \frac{\sigma^2_n - \bar n}{2\pi\, \bar n} +  2 \sum_{m=1}^{\infty} \frac{ V_m^2}{2\pi \bar n} \, \, \cos(m \, \phi_\Delta) \\ \nonumber
&\equiv& \frac{\Delta \rho_A[0]}{\sqrt{\rho_{A,ref}}} +   2\sum_{m=1}^{\infty}  \frac{\Delta \rho_A[m]}{\sqrt{\rho_{A,ref}}}\, \cos(m \, \phi_\Delta),
\eea
The first term $\Delta \rho_A[0] / \sqrt{\rho_{A,ref}}$ is the density ratio averaged over acceptance $(\Delta \eta,\Delta \phi)$. Its integral at full acceptance is normalized variance difference $\Delta \sigma_{n/}^2 \equiv (\sigma^2_n - \bar n)/\bar n$, which we divide by acceptance integral $1 \times 2\pi$ to obtain the mean of the 2D autocorrelation density ${\Delta \rho_A}[0]/{\sqrt{\rho_{A,ref}}} = {\Delta \sigma^2_{n/}(\Delta \eta, \Delta \phi)}/{\Delta \eta\, \Delta \phi} $. The sinusoid amplitudes are $ {\Delta \rho_A}[m]/{\sqrt{\rho_{A,ref}}}  = \overline{n(n-1) \,  \tilde v_m^2}/({2\pi} \bar n) \equiv \rho_0 \, v_m^2$, defining unbiased $v_m$. The {\em event-wise} $\tilde v_m^2 \equiv \langle \cos(m \phi_\Delta) \rangle$ are related to conventional flow measures $v_m$, but may be numerically quite different for small multiplicities due to bias in the latter. Different mean-value definitions result in different measured quantities ({\em cf.} App.~\ref{stats}). Whereas the $Q_m^2$ are power-spectrum elements which {\em include the white-noise reference}, the $V_m^2$ ($\propto$ squares of cylindrical multipole moments) represent the true azimuth correlation signal. This important result combines several measures of fluctuations and correlations within a comprehensive system.

\subsection{Azimuth vectors}

Power-spectrum elements $Q_m^2$ are derived from complex Fourier amplitudes ${\bf Q}_m$. In an alternate representation the complex ${\bf Q}_m$ can be replaced by {\em azimuth vectors} $\vec Q_m$. The  $\vec Q_m$ are conventionally referred to as  ``flow vectors''~\cite{poskvol}, but they include a statistical reference as well as a ``flow'' sinusoid. The $\vec V_m$ defined in this paper are more properly termed ``flow vectors,'' to the extent that such terminology is appropriate. We refer to the $\vec Q_m$ by the model-neutral term {azimuth} vector and define them by the following argument. 

The cosine of an angle difference---$\cos(m[\phi_1 - \phi_2])= \cos(m\phi_1)\cos(m\phi_2) + \sin(m\phi_1)\sin(m\phi_2)$---can be represented in two ways, with complex unit vectors ${\bf u}(m\phi) \equiv \exp(i\,m \phi)$ or with real unit vectors $\vec u(m\phi) \equiv  \cos(m\phi)\hat \imath +  \sin(m\phi)\hat \jmath$ [complex plane $(\Re z,\Im z)$ {\em vs} real plane $(x,y)$]. Thus,
 \bea
 \cos(m[\phi_1 - \phi_2]) &=& \Re\{ {\bf u}(m\phi_1)\,{\bf u}^*(m\phi_2)\}   \\ \nonumber
&=& \vec u(m\phi_1) \cdot \vec u(m\phi_2)). 
 \eea
If an analysis is reducible to terms in $\cos(m[\phi_1 - \phi_2])$  the same results are obtained with either representation. 
Thus, we can rewrite the first line of Eq.~(\ref{fourier})  as
\bea
\tilde \rho(\phi) &=&  \sum_{m = -\infty}^{\infty} \frac{\tilde{\vec Q}_m}{2\pi} \cdot \vec u(m\phi) \\ \nonumber
 &=&   \frac{\tilde Q_0}{2\pi} +   2\sum_{m = 1}^{\infty} \frac{\tilde Q_m}{2\pi} \cos( m [\phi - \Psi_m]),
\eea
in which case
\bea
{\tilde {\vec Q}_m} &=&  \int_{-\pi}^\pi d\phi \, \tilde \rho(\phi)\, \vec u(m\phi) = \sum_{i=1}^n r_i  \vec u(m \phi_i) \\ \nonumber
&=& n\langle r [\cos(m\phi),\sin(m\phi)] \rangle \\ \nonumber
&=&  \tilde Q_m \vec u(m \Psi_m),
 \eea
an event-wise random real vector.

\section{Conventional methods}

We now use the formalism in Sec.~\ref{foorier} to review conventional flow analysis methods in a common framework. We consider five significant papers in chronological order. The measurement of angular correlations to detect collective dynamics (e.g., parton fragmentation and/or hydrodynamic flow) proceeds from directivity (1983) at the Bevalac (1 - 2 GeV/u fixed target) to transverse sphericity predictions (1992) for the SPS/RHIC ($\sqrt{s_{NN}}$ = 17 - 200 GeV), then to Fourier analysis of azimuth distributions  (1994, 1998) and $v_2$ centrality trends (1999). We pay special attention to the manipulation of random variables (RVs). RVs do not follow the algebra of ordinary variables, and the differences are especially important for small sample numbers (multiplicities, {\em cf.} App.~\ref{stats}).
 
\subsection{Directivity at the Bevalac}

An important goal of the Bevalac HI program was observation of the {\em collective response} of projectile nucleons to compression during nucleus-nucleus collisions, called {\em directed flow}, which might indicate system memory of the initial impact parameter as opposed to isotropic thermal emission. Because of finite (small) multiplicities and large fluctuations relative to the measured quantity the geometry of a given collision may be poorly defined, but the final state may still contain nontrivial collective information. The analysis goal becomes separating possible collective signals from statistical noise. 

An initial search for collective effects was based on the 3D {\em sphericity} tensor $\tilde {\cal S} = \sum_i^n \vec p_i \vec p_i$~\cite{spherpart,spherflow} described in App.~\ref{sphere}.  Alternatively, the {\em directivity} vector was defined in the  transverse plane~\cite{danod}. In the notation of Sec.~\ref{foorier}, including weights $r_i \rightarrow  w_i p_{ti}$, directivity is
 \bea
 \tilde {\vec Q}_1 &\equiv& \sum_i^n w_i \vec p_{ti} = \sum_i^n w_i p_{ti} \vec u(\phi_i) \\ \nonumber
&=& \tilde Q_1 \vec u(\Psi_1),
 \eea
azimuth vector $\tilde {\vec Q}_m $ with m = 1 (corresponding to {\em directed} flow). Event-plane (EP) angle $\Psi_1$ estimates true reaction-plane (RP) angle $\Psi_r$. To maintain correspondence with SPS and RHIC analysis we simplify the description in~\cite{danod} to the case that $n$ single nucleons are detected and no  multi-nucleon clusters; thus $a \rightarrow 1$ and $A \rightarrow n$.

\begin{figure}[h]
  \includegraphics[width=1.65in,height=1.68in]{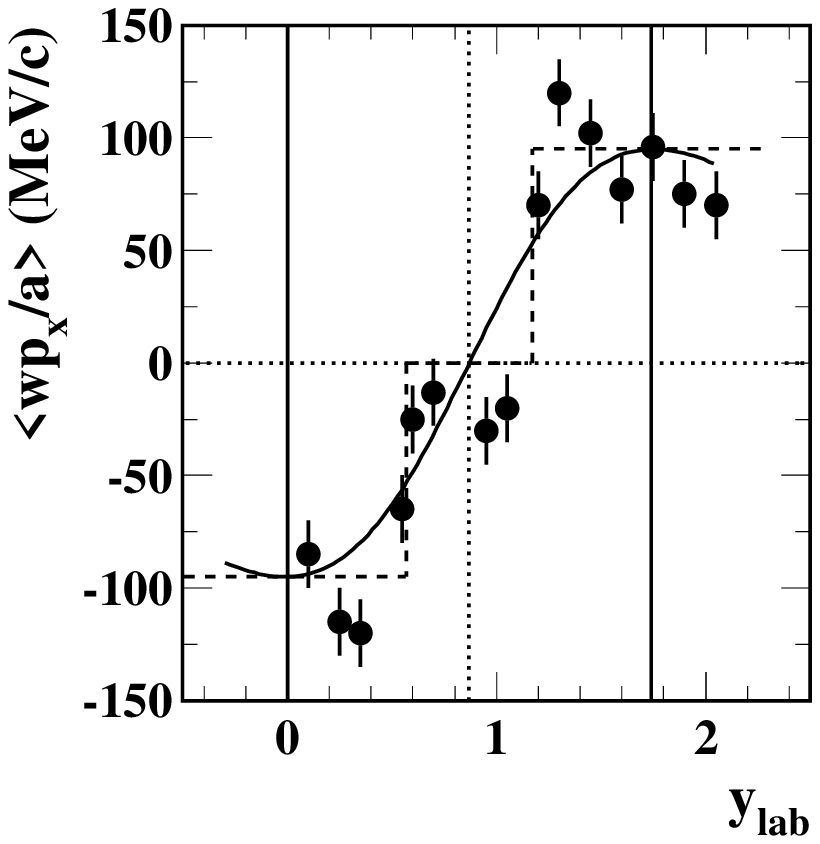}
\includegraphics[width=1.65in,height=1.68in]{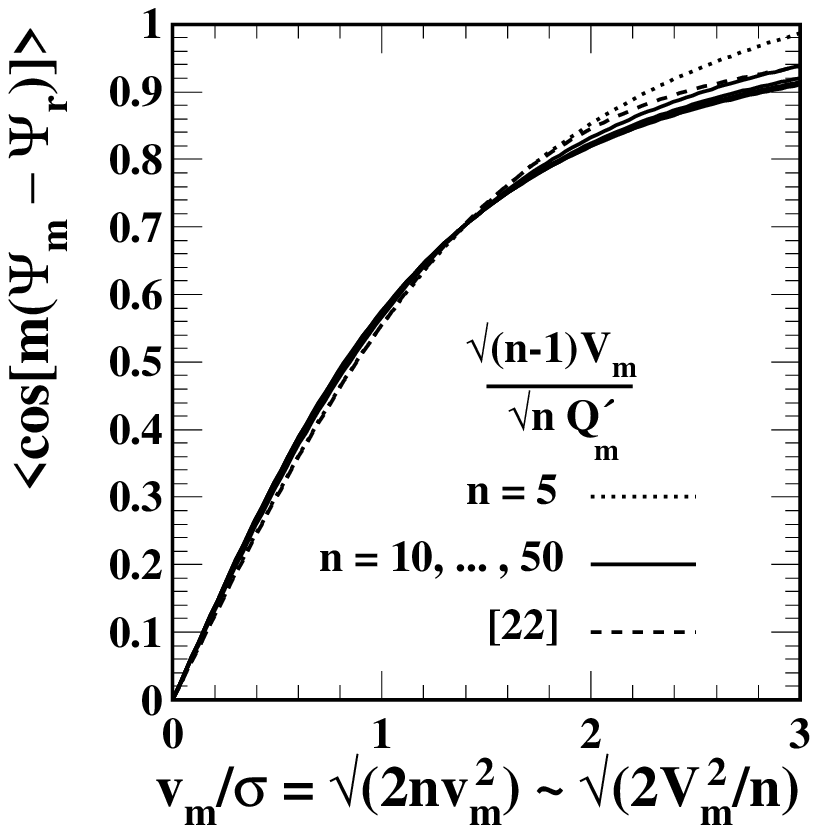}
 \caption{\label{fig1a}
Left panel: Comparison of directed flow data from Fig. 3(a) of the event-plane analysis and the $\Re Y_2^1 \propto \sin(2\theta[y_{lab}])$ spherical harmonic, with amplitude 95 MeV/c obtained from ${V_1^2}$. Weights in the form $w(y_{lab})$ are denoted by dashed lines. The correspondence with $\sin(2 \theta[y_{lab}])$ (solid curve) is apparent. Right panel: The EP resolution obtained from~\cite{poskvol} (dashed curve) and from ratio $\sqrt{(n-1)V^2_1 / nQ'^2_1}$ defined in this paper (solid and dotted curves for several $n$ values). 
 } 
 \end{figure}

The terms in $\tilde {\vec Q}_1$ are weighted by $w_i = w(y_i) = \pm 1$ corresponding to the forward or backward hemisphere relative to the CM rapidity, with a region $[-\delta,\delta]$ about mid-rapidity excluded from the sum ($w_i = 0$). $\tilde Q_1$ then approximates quadrupole moment $\tilde {\bf q}_{21}$ derived from spherical harmonic $\Re Y_2^1 \propto \sin(2\theta) \cos(\phi) $, as illustrated in Fig.~\ref{fig1a} (left panel, dashed lines). In effect, a rotated quadrupole is modeled by two opposed dipoles, point-symmetric about the CM in the reaction plane. It is initially assumed that EP angle $\Psi_1$ of vector $\vec Q_1$ estimates RP angle $\Psi_r$, and magnitude $Q_1$ measures directed flow.


The first part of the analysis was based on subevents---nominally equivalent but independent parts of each event. The dot product $\vec Q_{1A} \cdot \vec Q_{1B}$ for subevents A and B of each event was used to establish the existence of a significant flow phenomenon and the {\em angular resolution} of the RP estimation {\em via} the  distribution on $\Psi_{1A} - \Psi_{1B}$. The EP resolution was defined by $\overline{ \cos(\Psi_1 - \Psi_r) } = 2\sqrt{\overline{ \cos(\Psi_{1A} - \Psi_{1B}) }}$.  
The magnitude of $\tilde {\vec Q}_1$ was then related to an estimate of the mean transverse momentum in the RP. 
Integrating over rapidity with weights $w_i$ we obtain event-wise quantities
 \bea
 \tilde Q^2_1 &=& \sum_i^n w_i^2\, p_{ti}^2 + \sum_{i \neq j}^{n,n-1} w_i\, w_j\, \vec p_{ti} \cdot  \vec p_{tj} \\ \nonumber
&\simeq& n\langle p_t^2 \rangle + n(n-1) \langle p_t^2 \cos(\phi_\Delta) \rangle \\ \nonumber
&\equiv& \tilde Q_{ref}^2 + \tilde V_1^2.
 \eea
The last line makes the correspondence with the notation of this paper. 

The initial analysis in~\cite{danod} used $\tilde Q_1^2 - \tilde Q_{ref}^2 = \tilde V_1^2 =   n(n-1) \langle p^2_x \rangle$, assuming that $\hat x$ is contained in the RP and there are no non-flow correlations. Note that $n(n-1) \equiv \sum_{i \neq j}^{n,n-1} |w_i w_j|$ contains weights $w_i w_j$ implicitly. 
Since $w_i \sim \sin[2\, \theta(y_i)]$ ({\em cf.} Fig.~\ref{fig1a} -- left panel), what is actually calculated in~\cite{danod} for  the single-particle (no multi-nucleon clusters) case is the $p_t$-weighted r.m.s. mean of the spherical harmonic $\Re Y_2^1(\theta,\phi) \propto$ quadrupole moment $\tilde {\bf q}_{21}$, thereby connecting rank-1 tensor $\vec Q_1$ (with weights on rapidity) to rank-two sphericity tensor ${\cal S}$ ({\em cf.} App.~\ref{sphere}). 
The mean-square quantity calculated is
\bea \label{hatpx}
{ p^2_x } \equiv  {\frac{{V_1^2}}{\overline{n(n-1)}}} \simeq  { \frac{\overline{  \sin^2(2 \theta) p_t^2\cos^2(\phi - \Psi_r) }}{\overline{\sin^2(2 \theta)}}},
\eea
a minimally-biased statistical measure as discussed in App.~\ref{stats}, from which we obtain $p_x = \sqrt{p_x^2}$ estimating the transverse momentum per particle in the reaction plane.


The second part of the analysis sought to obtain ${ p_x(y) }$, the weighted-mean transverse momentum in the RP as a function of rapidity. It was decided to determine $ p_{xi} = p_{ti} \cos(\phi_i - \Psi_r)$ for the $i^{th}$ particle relative to the RP, but with $\Psi_r$ estimated by EP angle $\Psi_1$. The initial attempt was based on
\bea
w_i  \tilde p'_{xi} &\equiv& w_i \vec p_{ti} \cdot \vec u(\Psi_1) = w_i\vec p_{ti} \cdot \frac{\sum_j^n w_j\, \vec p_{tj}}{|\sum_k^n w_k\, \vec p_{tk}|}.
\eea
Summing $w_i\, \vec p_{ti}$ over all particles in a $y$ bin gives
\bea
 {\langle   p'_{x} \rangle} &=& \frac{\sum_i^n w^2_i\, p_{ti}^2 + \sum_{i \neq j}^{n, n-1}  w_i w_j\, \vec p_{ti} \cdot \vec p_{tj}}{\sum_l^n |w_l| \,|\sum_k^n w_k \,\vec p_{tk}|} \\ \nonumber
& = & \frac{n  \langle p_t^2\rangle  + \tilde V_1^2}{n\, \tilde Q_1}    \\ \nonumber
&=&  \tilde Q_1/ n = \sqrt{\frac{1}{n}\langle   p^2_{t} \rangle + \frac{n-1}{n}\langle   p^2_{x} \rangle},
\eea
from which we obtain ensemble mean $p'_x = \overline{\tilde {\langle p'_x\rangle}}$.
That result can be compared directly with the linear speed inferred from a random walker trajectory, which is `infinite' in the limit of zero time interval ({\em cf.} App.~\ref{brown}). The first term in the radicand is said to produce ``multiplicity distortions''  in conventional flow terminology. The second term contains the unbiased quantity. 

In contrast to the first part of the analysis the second method retains the statistical reference within $\tilde Q_1$ as part of the result, so that $\tilde Q_1 / n \sim \sqrt{ \langle p^2_t \rangle /n} $ for small multiplicities and/or flow magnitudes, a false signal comparable in magnitude to the true flow signal. The unbiased directed flow $ p_x $ was said to be ``distorted'' by the presence of the statistical reference (called unwanted self-correlations or ``autocorrelations'') to the strongly-biased value $ p'_x $ dominated by the statistical reference.

An attempt was made to remove statistical distortions arising from self pairs by redefining $\tilde {\vec Q}_1 \rightarrow \tilde {\vec Q}_{1i}$, a vector complementary to each particle $i$ with that particle omitted from the sum. The estimator of $\Psi_r$ for particle $i$ is then $\Psi_{1i}$ in $\tilde {\vec Q}_{1i} \equiv  \sum_{j \neq i}^{n-1}w_j\, \vec p_{tj} = \tilde  Q_{1i}\, \vec u(\Psi_{1i})$ and
\bea
w_i \tilde p''_{xi} &=& w_i\vec p_{ti} \cdot \vec u(\Psi_{1i}) = w_i\vec p_{ti} \cdot \frac{\sum_{j \neq i}^{n-1} w_j\, \vec p_{tj}}{|\sum_{k \neq i}^{n-1} w_k\, \vec p_{tk}|}.
\eea
Summing over $i$ within a rapidity bin one has
\bea \label{wrong}
 {\langle   p''_{x} \rangle}  &=& \frac{1}{\sum_l^n |w_l|} \sum_i^n \frac{ \sum_{j \neq i}^{n-1} w_i w_j\, \vec p_{ti} \cdot \vec p_{tj}}{ |\sum_{k \neq i}^{n-1} w_k\, \vec p_{tk}|} \\ \nonumber
&\simeq&  \frac{\tilde V_1^2}{ n\tilde Q'_{1}} \\ \nonumber
&=&\sqrt{\langle   p^2_{x} \rangle}\, \frac{\sqrt{(n-1)\langle   p^2_{x} \rangle}}{\sqrt{\langle   p^2_{t} \rangle + (n-2)\langle   p^2_{x} \rangle}} \\ \nonumber
&\simeq& \sqrt{\langle   p^2_{x} \rangle} \, \langle \cos(\Psi'_m - \Psi_r) \rangle
\eea
with $\tilde Q'_{1} \equiv  \sqrt{(n-1)\langle p_t^2 \rangle + (n-1)(n-2) \langle p_t^2 \cos(\phi_\Delta)\rangle}$. Since ${V_1^2} = \overline{n(n-1)  \langle p_x^2 \rangle} =   \overline{n(n-1) \langle p_t^2 \cos(\phi_\Delta)\rangle} $ one sees that the division by $\tilde Q'_{1}$ is incorrect, even though it seems to follow the chain of argument based on RP estimation and EP resolution with correction. The correct (minimally-biased) quantity is $p_x \equiv \sqrt{\overline{p^2_x}} = \sqrt{V_1^2 / \overline{n(n-1)}}$. The new EP definition removes the reference term from the numerator, but $\tilde Q'_1$ in the denominator retains the statistical reference in $ p''_x $.  There is the additional issue that $\sqrt{\overline{x^2}} \neq \bar x$. Two different mean values are represented by $\overline{\langle p_x \rangle}$ and $p_x = \sqrt{\overline{p_x^2}}$. The difference can be large for small event multiplicities.

The remaining bias was attributed to the EP resolution. The resolution correction factor derived from the initial subevent analysis ({\em cf.} App.~\ref{subev}) was applied to $\langle p''_x \rangle$ to further reduce bias. In Fig.~\ref{fig1a} (right panel) we compare the EP resolution correction from~\cite{poskvol} (dashed curve) with factor  $\sqrt{\overline{ (n-1) \tilde V^2_1} / \overline{ n\tilde Q'^2_1}} $ required to convert $\overline{\langle p''_x \rangle}$ from Eq.~(\ref{wrong}) to $p_x$ from Eq.~(\ref{hatpx}). The agreement is very good. Eq. (\ref{hatpx}) is the least biased and most direct way to obtain ${p_x} \equiv \sqrt{\overline{p^2_x}}$, both globally over the detector acceptance and locally in rapidity bins, without EP determination or resolution corrections. 

In Fig.~\ref{fig1a} (left panel) we show the data (points) for $\overline{ p_x }(y_{lab})$ from the EP-corrected analysis and the solid curve $p_x  \, \sin[2 \theta(y_{lab})] \propto   {\bf q}_{21} Y_{21}(\theta(y),0)$, where $p_x  = \sqrt{V_1^2/\overline{n(n-1)}} = 95$ MeV/c. The agreement is good, and the similarity of $\sin[2 \theta(y_{lab})]$ (solid curve) to weights $w(y_{lab})$ (dashed lines) noted above is apparent. Location of the $\sin[2 \theta(y_{lab})]$ extrema near the kinematic limits (vertical lines) is an accident of the collision energy and nucleon mass. 
These results are for $E_{cm} =1.32~\text{GeV} \sim \sqrt{2} m_N c^2$. 
In the notation of this paper  ${V_1^2} = 4.7$ (GeV/c)$^2$, $\sqrt{{V_1^2} /\overline{n(n-1)}} = 0.095$ GeV/c = $\overline{w p_x/a}$, and $\overline{Q_x} \equiv \bar n \sqrt{{V_1^2} /\overline{n(n-1)}} = 2.17~ \text{GeV/c}~ \simeq \sqrt{V_1^2}$ (not $Q_1$). By direct and indirect means (directivity and RP estimation) quadrupole moment ${\bf q}_{21} \propto V_1$ was measured.

\subsection{Transverse sphericity at higher energies}

The arguments and techniques in~\cite{ollitrault} suggest a smooth transition from relativistic Bevalac and AGS energies (collective nucleon and resonance flow) to intermediate SPS and ultra-relativistic  RHIC energies (possible  transverse flow, possibly QCD matter, collectivity manifested by correlations of produced hadrons, mainly pions). For all heavy ion collisions thermalization is a key issue. Clear evidence of thermalization is sought, and collective flow is expected to provide that evidence. 

Two limiting cases are presented in~\cite{ollitrault} for SPS flow measurements: 1) linear N-N superposition with no collective behavior (no flow); 2) thermal equilibrium -- collective pressure in the reaction plane -- fluid dynamics leading to ``elliptic'' flow. In a hydro scenario the initial space eccentricity transitions to momentum eccentricity through thermalization and early pressure. The paper considers flow measurement techniques appropriate for the SPS and RHIC, and in particular newly defines transverse sphericity ${\cal S}_t$.

According to~\cite{ollitrault} the 3D sphericity tensor introduced at lower energies~\cite{spherflow} can be simplified in ultra-relativistic heavy ion collisions to a 2D {transverse} sphericity tensor. Sphericity is transformed to 2D by $\vec p \rightarrow \vec p_t$, omitting the momentum $\hat z$ component near mid-rapidity. Transverse sphericity (in dyadic notation) is
 \bea \label{transpher}
 2\tilde {\cal S}_t &\equiv& 2\sum_i^n \vec p_{ti} \vec p_{ti} \\ \nonumber
&=& \sum_i p_{ti}^2 \left\{  {\cal I} +  {\cal C}(\phi_i) \right\} \\ \nonumber
&\equiv& n \langle p_{t}^2 \rangle \left \{   {\cal I} + \tilde \alpha_1\,   {\cal C}(\Psi_2) \right\} 
 \eea
defining $\tilde \alpha_1$ and $\Psi_2$ in the tensor context, with 
\bea
 {\cal C}(\phi) \equiv \left[ \begin{array}{cc}
\cos(2\phi)&\sin(2\phi)\\
\sin(2\phi)&-\cos(2\phi)
\end{array} \right].
\eea
This $\tilde \alpha$ definition corresponds to Eq. (3.1) of ~\cite{ollitrault}. 

We next form the contraction of $\tilde {\cal S}_t$ with itself
\bea
2\tilde {\cal S}_t : \tilde {\cal S}_t   &=& 2\sum_{i,j}^{n,n} (\vec p_{ti}\cdot \vec p_{tj})^2 \\ \nonumber
&=& 2\sum_i p^4_{ti} + 2\sum_{i \neq j} p^2_{ti} p^2_{tj} \cos^2(\phi_i - \phi_j)  \\ \nonumber
&= & 2 n \langle p_t^4 \rangle + 2n(n-1) \langle p_t^4 \cos^2 (\phi_\Delta) \rangle \\ \nonumber
&= & n(n+1)\langle p_t^4 \rangle + n(n-1)  \langle p_t^4 \cos (2\phi_\Delta) \rangle 
\eea
using the dyadic contraction notation ${{\cal A}:{\cal B}}~\equiv {\cal A}_{ab}{\cal B}_{ab}$, with the usual summation convention. That self-contraction of a rank-2 tensor can be compared to the more familiar self-contraction of a rank-1 tensor $\tilde {\vec Q}_2 \cdot \tilde {\vec Q}_2 = \tilde Q_2^2 =  n \langle p_t^2 \rangle + n(n-1)\langle p_t^2 \cos(2\phi_\Delta) \rangle$. The quantity $2[\tilde {\cal S}_t : \tilde {\cal S}_t]_{ref} = n(n+1) \langle p_t^4 \rangle$  is the (uncorrelated) reference for the rank-2 contraction, whereas $\tilde Q^2_{2,ref} = n \langle p_t^2 \rangle$ is the reference for the rank-1 contraction. Subtracting the rank-2 reference contraction gives
\bea 
2 \left \{ \tilde {\cal S}_t : \tilde {\cal S}_t - [\tilde {\cal S}_t : \tilde {\cal S}_t]_{ref} \right\}
&= & n(n-1) \langle p_t^4 \cos(2\phi_\Delta) \rangle \nonumber \\
&\simeq&  \langle p_t^2 \rangle \tilde  V_2^2,
\eea
which relates transverse sphericity to two-particle correlations in the form $\tilde V_2^2 = \tilde Q_2^2 - \tilde Q_{2,ref}^2$, thus establishing the exact correspondence between $\tilde {\cal S}_t$ and $\tilde {\vec Q}_2$.

From the definition of $\alpha_1$ in Eq.~\ref{transpher} above and Eq. (3.1) of~\cite{ollitrault} we also have
\bea
2\,\tilde {\cal S}_t :\tilde {\cal S}_t  &=& n^2 \langle p_t^2 \rangle^2 (1 + \tilde \alpha_1^2)
\eea 
which implies
\bea \label{tensor}
n^2 \langle p_t^2 \rangle^2 \tilde \alpha_1^2 &= & n^2\, \tilde \sigma^2_{ p_t^2 } + n \langle p_t^4 \rangle \\ \nonumber
&+&  n(n-1) \langle p_t^4 \cos(2 \phi_\Delta) \rangle \nonumber \\
&\simeq& n^2\, \tilde \sigma^2_{ p_t^2 } +  \langle p_t^2\rangle \tilde Q_2^2.
\eea
The first relation is exact, given the definition of $\tilde \alpha_1$, but produces a complex statistical object containing the event-wise variance of $ p_t^2 $ in its numerator and random variable $n^2$ in its denominator. For $1/n \rightarrow 0$ it is true that $\tilde \alpha_1 \rightarrow \langle \cos(2[\phi - \Psi_r])\rangle$ (since $\Psi_2 \rightarrow \Psi_r$ also), but $\tilde \alpha_1$ is a strongly biased statistic for finite $n$. 

The definition  
 \bea
 \tilde\alpha_2 = \frac{\langle p_x^2 - p_y^2 \rangle}{\langle p_x^2 + p_y^2 \rangle} 
 \eea
from Eq. (2.5) of~\cite{ollitrault} seems to imply $\tilde \alpha_2 = \langle p_t^2 \cos(2[\phi - \Psi_r])\rangle / \langle p_t^2 \rangle \rightarrow \langle \cos(2[\phi - \Psi_r]) \rangle$, assuming that $\hat x$ lies in the RP. However, the latter relation fails for finite multiplicity [the effect of the statistical reference or self-pair term $n$ in Eq.~(\ref{tensor})] because each of event-wise $\langle p_x^2 \rangle$ and $\langle p_y^2 \rangle$ is a random variable, and their independent random variations do not cancel in the numerator. The exact relation is the first line of
\bea \label{qm}
n^2 \langle p_t^2 \rangle^2 \tilde \alpha_2^2 &=& n^2\langle p_t^2 \cos(2[\phi - \Psi_2]) \rangle^2 \\ \nonumber
&\simeq&  n^2 \langle p_t^2 \rangle \langle p_t \cos(2[\phi - \Psi_2]) \rangle^2 \\ \nonumber
& = & \langle p_t^2 \rangle\,  \tilde Q_2^2 \\ \nonumber
\tilde \alpha_2 &\simeq& \tilde Q_2 / \tilde Q_0 \neq \tilde V_2 / \tilde Q_0
\eea
The second line is an approximation which indicates that $\tilde \alpha_2$ is more directly related to $\tilde Q_2$ than is $\tilde \alpha_1$. But $ Q_2^2$ is a poor substitute for $ V_2^2$ which represents true two-particle azimuth correlations in a minimally-biased way by incorporating a proper statistical reference. The effect of the reference contribution is termed a `distortion' in~\cite{ollitrault}.

\subsection{Fourier series I} \label{fourier1}

Application of Fourier series to azimuth particle distributions was introduced in~\cite{volzhang}. Fourier analysis is described as model independent, providing variables which are ``easy to work with and have clear physical interpretations.'' Sinusoids or harmonics are associated with transverse collective flow, the model-dependent  language following~\cite{ollitrault} closely. To facilitate comparisons we convert notation in~\cite{volzhang} to that used in this paper: $r(\phi) \rightarrow \rho(\phi)$, $(x_m,y_m) \rightarrow  {\vec Q}_m$, $\psi_m \rightarrow \Psi_m$, $v_m \rightarrow  Q_m$ and $\tilde v_m \rightarrow V_m$. 

According to the proposed method density $\tilde \rho(\phi)$ represents, within some $(p_t,\eta)$ acceptance, an event-wise particle distribution on azimuth $\phi$ including weights $r_i = 1,~ p_{ti}$ or $E_{ti}$. The FT in terms of azimuth vectors is
\bea
 \tilde \rho(\phi) &=& \sum_i^n r_i \delta(\phi - \phi_i) \\ \nonumber
  &=& \frac{\tilde Q_0 }{2 \pi} +  2\sum_{m=1}^{\infty} \frac{\tilde {\vec Q}_m}{2\pi} \cdot \vec u(m\phi) \\ \nonumber 
&=& \frac{\tilde Q_0 }{2 \pi} + 2 \sum_{m=1}^{\infty} \frac{\tilde { Q}_m}{2\pi} \cos(m[\phi - \Psi_{m}]) ,
\eea
and the RT is
\bea
\tilde {\vec Q}_m &=& \sum_i^n r_i \vec u(m \phi_i) \equiv \tilde Q_m \vec u(m \Psi_m),
\eea
forming a conventional Fourier transform pair [{\em cf.} Eqs.~(\ref{fourier}) and (\ref{freverse})]. Scalar amplitude $\tilde Q_m = \sum_{i}^n r_i \vec u(m \phi_i) \cdot \vec u(m \Psi_m) = n \langle r \cos(m[\phi - \Psi_m])\rangle  $ is the proposed flow-analysis quantity. $\tilde Q_m$ is said to measure the flow magnitude, and $\Psi_m$ estimates reaction-plane angle $\Psi_r$. It is proposed that $\tilde Q_m(\eta)$ evaluated within bins on $\eta$ may characterize complex ``event shapes'' [densities on $(\eta,\phi)$].

As with directivity and sphericity, multiplicity fluctuations are seen as a major obstacle to flow analysis with Fourier series. Finite multiplicity is described as a source of `bias' which must be suppressed. It is stated that $(\tilde V_m,\Psi_r)$ are the ``parameter[s] relevant to the magnitude of flow,'' whereas the observed $(\tilde Q_m,\Psi_m)$ are biased flow estimators. In the limit $1/n \rightarrow 0$ the two cases would be identical. A requirement is therefore placed on minimum event-wise multiplicity in a ``rapidity slice.'' To solve the finite-number problem the paper proposes to use the event frequency distribution on $\tilde Q^2_m$ from event-wise Fourier analysis to measure flow. 

If correlations are  zero then
 \bea
 \tilde Q_m^2 \rightarrow \tilde Q^2_{ref} =  \sum_i^n r_i^2 = n \langle r^2 \rangle \equiv \tilde \sigma^2 ,
 \eea
with $\overline{\langle r^2 \rangle} \equiv \sigma^2_0$.  By fitting the frequency distribution on $\tilde Q_m^2$ with a model function it is proposed to obtain $\tilde V_m$ as the unbiased flow estimator. The fitting procedure is said to require sufficiently large event multiplicities to obtain $\tilde V_m$ unambiguously. 

\begin{figure}[h]
  \includegraphics[width=1.65in,height=1.68in]{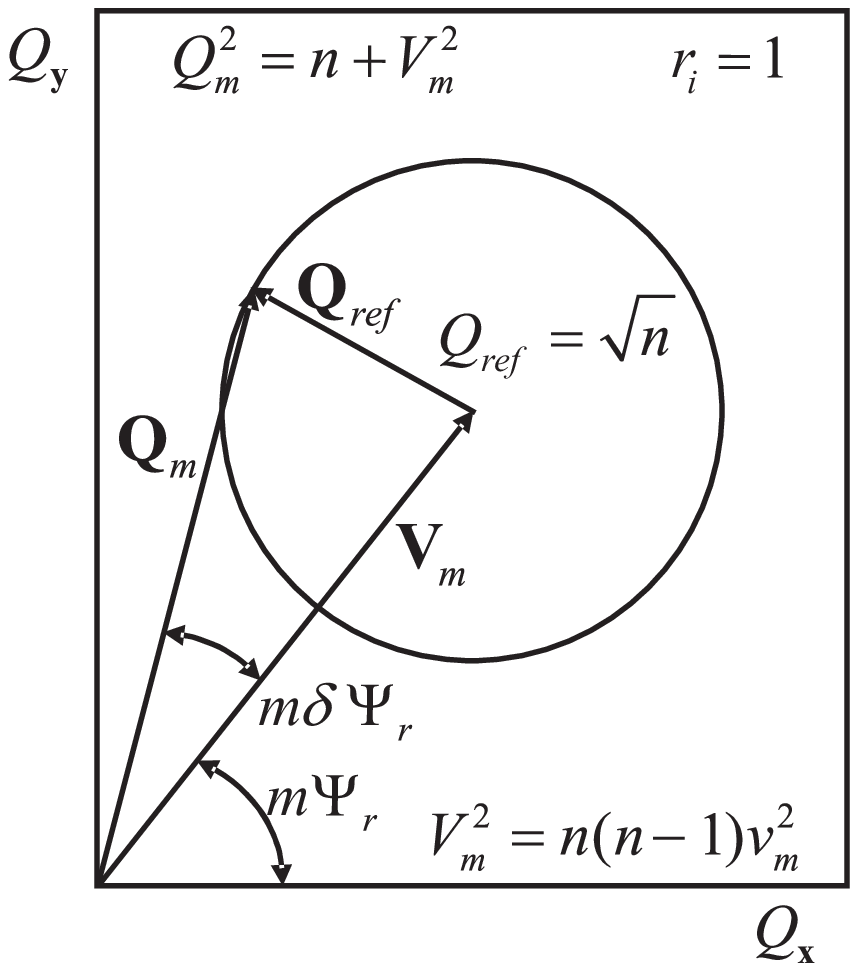}
\includegraphics[width=1.65in,height=1.68in]{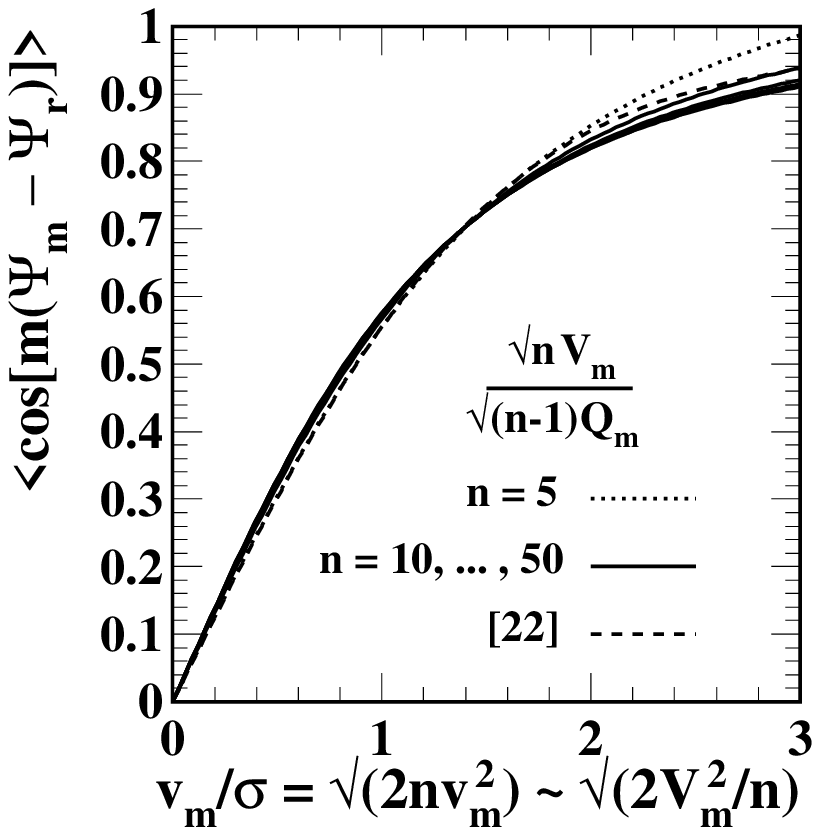}
 \caption{\label{fig1z}
 Left panel: Distribution of event-wise elements of  $\tilde {\vec Q}_m$ components determined by the gaussian-distributed random walker $\tilde {\vec Q}_{ref}$ and possible correlation component $\tilde {\vec V}_m$. Right panel: Reaction-plane resolution estimator $\langle \cos(m\delta \Psi_{mr}) \rangle$, with $\delta \Psi_{mr} = \Psi_m - \Psi_r$, determined from fits to a distribution on $\tilde Q_m^2$ as in the left panel (dashed curve), and from Eq.~\ref{epres} for several values of $n$ (solid and dotted curves).  
 } 
 \end{figure}

The distribution on $\tilde Q_m^2$ is derived as follows ({\em cf.} Fig.~\ref{fig1z} -- left panel). The magnitude of statistical reference $\tilde {\vec Q}_{ref}$ (a random walker) has probability distribution $ \propto \exp(-\tilde Q_{ref}^2/2 Q_{ref}^2)$, with $Q_{ref}^2 = \overline{ n \langle r^2 \rangle }$. But $\tilde {\vec Q}_{ref} = \tilde {\vec Q}_m - \tilde {\vec V}_m$, therefore ({\em cf.} Fig.~\ref{fig1z} -- left panel) 
\bea \label{hardway}
\tilde Q_{ref}^2 &=& \tilde Q_m^2 + \tilde V_m^2 - 2 \tilde { Q}_m  \tilde { V}_m \cos(m [\Psi_m - \Psi_r]),
\eea
and 
\bea
 \exp(-\tilde Q_{ref}^2/2 Q_{ref}^2) \rightarrow \rho(\tilde Q_m,\Psi_m; \tilde V_m,\Psi_r).
\eea 
When integrated over $\cos(m [\Psi_m - \Psi_r])$ there results the required probability distribution on $\tilde Q_m^2 $, with fit parameter $V_m$. 
The distribution on $\tilde Q^2_m$ is said to show a `nonstatistical' shape change from which $ V_m$ can be inferred by a model fit  ``{\em free from uncertainties} in event-wise determination of the reaction plane.''  It is also proposed to use $\rho(\tilde Q_m,\Psi_m;\tilde V_m,\Psi_r)$ to determine the EP resolution $\overline{ \cos(\Psi_m - \Psi_r)} $ by integrating over $\tilde Q^2_m$ and using the resulting projection on $\cos(\Psi_m - \Psi_r)$ to determine the ensemble mean (Fig. 3 of~\cite{volzhang}). 

While one could extract ensemble-average $ V^2_m$ at some level of accuracy by fitting the frequency distribution on $\tilde Q^2_m$ with a model function, we ask {\em why go to that trouble} when  $V^2_m$ is easily obtained as a variance difference? Instead of Eq.~(\ref{hardway}) we simply write
\bea
\tilde Q_m^2 &=&  \tilde Q_{ref}^2 + \tilde V_m^2,
\eea
where the cross term is zero on average and $\tilde Q^2_{ref} = \ n \langle r^2 \rangle$ represents the power-spectrum white noise, which is the same for all $m$ (i.e., `white'). If the vector mean values are zero that is a relation among variances. Ensemble mean $V^2_m = Q_m^2 - Q_{ref}^2$ is therefore simply determined.

For the EP resolution we factor $\tilde V_m^2$
\bea
\tilde V_m^2 &=& n(n-1) \langle r^2\, \cos^2(m[\phi - \Psi_r]) \rangle \\ \nonumber
&=& n(n-1) \langle r^2\, \cos^2(m[\phi - \Psi_m])\rangle \,   \cos^2(m[\Psi_m - \Psi_r]). 
\eea
We use $ \tilde Q_m =  n \langle r \cos(m[\phi - \Psi_m])\rangle$ and the assumption that random variable $\Psi_m - \Psi_r$ is uncorrelated with $\phi - \Psi_m$ to obtain
\bea \label{epres}
\overline{ \cos^2(m[\Psi_m - \Psi_r])} =   \frac{ \overline{n\,\tilde V_m^2}}{\overline{ (n-1)\, \tilde Q_m^2}},
\eea
which defines the EP resolution of the full $n$-particle event in terms of power-spectrum elements ({\em cf.} App.~\ref{subev}). The square root of that expression is plotted in Fig.~\ref{fig1z} (right panel) as the solid curves for several values of $\bar n$. The solid and dotted curves are nearly identical to those in Fig.~\ref{fig1a}.

As in other flow papers there is much emphasis on insuring adequate multiplicities to reduce bias to a managable level, because an easily-determined statistical reference is not properly subtracted to reveal the contribution from true two-particle correlations in isolation. For $\sqrt{2n}\, v_m > 1$ in Fig.~\ref{fig1z} (right panel) the EP is meaningful; bias of event-wise quantities relative to the EP is manageable. For $\sqrt{2n}\, v_m < 1/2$ the EP is poorly defined, and ensemble-averaged two-particle correlations are the only reliable measure of azimuth correlations. In either case EP estimation is only justified when a non-flow phenomenon is to be studied relative to the reaction plane.

\subsection{Fourier series II} \label{fourier2}


A more elaborate review of flow analysis methods based on Fourier series is presented in~\cite{poskvol}. The approach is said to be general. The event plane is obtained for each event.
The event-wise Fourier amplitude(s) $\tilde q_m \equiv \tilde Q_m / \tilde Q_0$ relative to the EP are corrected for the EP resolution as obtained from subevents. We modify the notation of the paper to $v_m^{obs} \rightarrow \tilde q_m$ and $w_i \rightarrow r_i$  to maintain consistency within this paper. We distinguish between the unbiased $\tilde v_m \equiv \tilde V_m / \tilde V_0$ 
and the biased $\tilde q_m$.

According to~\cite{poskvol}, in the $1/n \rightarrow 0$ limit ($\tilde Q_m \rightarrow  V_m$, no tildes, no random variables) the dependence of the single-particle density on azimuth angle $\phi$ integrated over some ($p_t,y$) acceptance can be expressed as a Fourier series of the form
 \bea
  \rho(\phi) = \frac{ V_0}{2\pi} \left\{ 1 + 2\sum_{m=1}^\infty  v_m \cos\left[m(\phi - \Psi_r)\right] \right\},
 \eea
with reaction-plane angle $\Psi_r$. As we have seen, the factor 2 comes from the symmetry on index $m$ for a real-number density, not an arbitrary choice as suggested in~\cite{poskvol}. In this definition $ V_0 / 2\pi$ has been factored from the Fourier series in~\cite{volzhang}. The Fourier ``coefficients'' $ v_m$ in this form (actually coefficient ratios) are not easily related to the power spectrum.  In the $1/n \rightarrow 0$ limit the coefficients are $ v_m = \langle \cos(m[\phi - \Psi_r])\rangle$. 

In the analysis of finite-multiplicity events, reaction-plane angle $\Psi_r$ defined by the collision (beam) axis and the collision impact parameter is estimated by event plane (EP) angle $\Psi_m$, with  $\Psi_m$ derived from event-wise {azimuth vector} $\tilde {\vec Q} _m$ (conventional flow vector)
 \bea
 \tilde {\vec Q}_m &=& \sum_{i=1}^n r_i \vec u(m\phi_i) \equiv \tilde Q_m\vec u(m\Psi_m).
 \eea
The finite-multiplicity event-wise FT is
\bea
\tilde  \rho(\phi) = \frac{\tilde Q_0}{2\pi} \left\{ 1 + 2\sum_{m=1}^\infty  \tilde q_m \cos\left[m(\phi - \Psi_m)\right] \right\}.
 \eea
with $\tilde q_m = \langle \cos(m[\phi - \Psi_m])\rangle$; e.g.,  $\tilde q_2 \simeq \tilde \alpha_2$ ({\em cf.} Eq.~(\ref{qm})). 

According to the conventional description the EP angle is biased by the presence of self pairs, unfortunately termed the ``autocorrelation effect'' or simply ``autocorrelation'' in conventional flow analysis~\cite{danod,poskvol}, whereas autocorrelations and cross-correlations are distributions on difference variables used for decades in statistical analysis to measure correlation structure on time and space. As in~\cite{danod}, to eliminate ``autocorrelations'' EP angle $\Psi_{mi}$ is estimated for each particle $i$ from complementary flow vector $\vec Q_{mi}= \sum_{j \neq i}^{n-1} r_j \vec u(2\phi_j) = Q_{mi} \vec u(m \Psi_{mi})$, a form of subevent analysis with one particle {\em vs} $n - 1$ particles ({\em cf.} App.~\ref{subev}).

 
In the conventional description the {event-plane resolution} results from fluctuations $\delta \Psi_r \equiv \Psi_m - \Psi_r$ of the event-plane angle $\Psi_m$ (or $\Psi_{mi}$) relative to the true reaction-plane angle $\Psi_r$ ({e.g.,} due to finite particle number). The EP resolution is said to reduce the observed $\tilde q_m$ relative to the true value $\tilde v_m$:
 \bea \label{littleq}
 \tilde q_{m}  &=&  \sqrt{\frac{1}{n} + \frac{n-1}{n} \tilde v_m^2} \\ \nonumber
&=& \langle \cos \left( m [\Psi_r - \Psi_m] \right) \rangle \cdot \tilde v_{m}.
 \eea
The EP resolution (first factor, second line) is obtained in a conventional flow analysis in two ways:  the frequency distribution on $\Psi_m - \Psi_r$ discussed in Sec.~\ref{fourier1} and the subevent method discussed in App.~\ref{subev}. A parameterization from the frequency-distribution method reported in~\cite{poskvol} is plotted as the dashed curve in Fig.~\ref{fig1z} (right panel). There is good agreement with the simple expression $\sqrt{n/(n-1)}\,V_m / Q_m $ obtained in Eq.~(\ref{epres}), which also follows from Eq.~(\ref{littleq}).


Two methods are described for obtaining $v_m$ without an event-plane estimate, with the proviso that large event multiplicities in the acceptance are required. The first, in terms of the conventional flow vector, is expressed (with $r_i \rightarrow 1$) as (Eq.~(26) of~\cite{poskvol})
\bea
Q_m^2 = \bar n + \overline{n^2} \, \bar {v}^2_m.
\eea
That expression is constrasted with the exact event-wise treatment, where for each event we can write
 \bea \label{quick}
\tilde  Q_m^2 &=& n + \sum_{i \neq j} \cos\left[ m (\phi_i - \phi_j) \right] \\ \nonumber
 &=&  n + n(n-1) \langle \cos ( m \phi_\Delta ) \rangle \\ \nonumber
 &\equiv& n + n(n-1) \tilde v_m^2 \\ \nonumber
\rightarrow Q_m^2 &=& \bar n + \overline{n(n-1) \tilde v_m^2} = \bar n + V_m^2
 \eea
Note the similarity with fluctuations measured by number variance $\sigma^2_n = \bar n + \Delta \sigma^2_n$, where the second term on the RHS is an integral over two-particle number correlations and the first is the uncorrelated Poisson reference (again, the self-pair term in the autocorrelation).

There are substantial differences between the two $Q_m^2$ formulations above, especially for smaller multiplicities. Eq.~(\ref{quick}) is unbaised for all $n$ and provides a simple way to obtain $ V_m^2 = \overline{ n(n-1) \tilde  v_m^2} = Q_m^2 - \bar n$. The conventional method uses a complex fit to the frequency distribution on $\tilde Q^2_m$ to estimate $ V_m$ as in~\cite{volzhang}. Why do that when such a simple alternative is available?

 \subsection{$v_2$ centrality dependence}

In~\cite{volposk} the expected trend of $v_2$ with A-A centrality for different collision systems is discussed in a hydro context. It is stated that the $v_2$ centrality trend should reveal the degree of equilibration in A-A collsions. The centrality dependence of $v_2 / \epsilon$ should be sensitive to the ``physics of the collision''---the nature of the constituents (hadrons or partons) and their degree of thermalization or collectivity. ``It is understood that such a state requires (at least local) thermalization of the system brought about by many rescatterings per particle during the system evolution...$v_2$ is an indicator of the degree of equilibration.''

Thermalization is related to the number of rescatterings, which also strongly affects elliptic flow according to this hydro interpretation. In the full hydro limit, corresponding to full thermalization where the mean free path $\lambda$ is much smaller than the flowing system, relation $v_2 \propto \epsilon$ is predicted, with $\epsilon$ the space eccentricity of the initial A-A overlap region~\cite{ollitrault}. Conversely, in the {\em low-density limit} (LDL), where $\lambda$ is comparable to or larger than the system size, a different model predicts the relation $v_2 \propto \epsilon\, A^{1/3}/\lambda$, where $A^{1/3}/\lambda$ estimates the mean number of collisions per ``particle'' during system evolution to kinetic decoupling~\cite{heisel}.  In the LDL case $v_2 \propto  \epsilon \frac{1}{S}\frac{dn}{dy}$, where  $S = \pi R_x R_y$ is the (weighted) cross-section area of the collision and $\epsilon = \frac{R_y^2 - R_x^2}{R_y^2 + R_x^2}$ is the spatial eccentricity. Those trends are further discussed in App.~\ref{centro}.

According to the combined scenario, comparison of the centrality dependence of $v_2$ at energies from AGS to RHIC may reveal a transition from hadronic to thermalized partonic matter.  The key expectation is that at some combination(s) of energy and centrality $v_2 / \epsilon$ transitions from an LDL trend (monotonic increase) to hydro (saturation), indicating (partonic) equilibration. 

However, it is important to note two things: 1) That overall description is contingent on the strict hydro scenario. If the quadrupole component of azimuth correlations arises from some other mechanism then the descriptions in~\cite{ollitrault} and~\cite{heisel} are invalid, and $v_2$ does not reveal the degree of thermalization. 2) $v_2$ is a model-dependent and statistically-biased quantity motivated by the hydro scenario itself. The model-independent measure of azimuth quadrupole structure is  $V_2^2 / \bar n \equiv  \bar n \, v_2^2$ (defining an unbiased $v_2$). It is important then to reconsider the azimuth quadrupole centrality and energy trends revealed by that measure to determine whether a hydro interpretation is a) required by or even b) permitted by data.

\section{Is the event plane necessary?}

A key element of conventional flow analysis is estimation of the reaction plane and the resolution of the estimate. We stated above that determination of the event plane is irrelevant if averaged quantities are extracted from an ensemble of event-wise estimates. The reaction-plane angle is relevant only for study of nonflow (minijet) structure relative to the reaction plane on $\phi_\Sigma$, the sum (pair mean azimuth) axis of $(\phi_1,\phi_2)$. In contrast to concerns about low multiplicities in conventional flow analysis, proper autocorrelation techniques accurately reveal ``flow'' correlations (sinusoids) and any other azimuth correlations, even in p-p collisions and even within small kinematic bins. In this section we examine the necessity of the event plane in more detail.

The reaction plane (RP), nominally defined by the beam axis and the impact parameter between centers of colliding nuclei, is determined statistically in each event by the distribution of participants. The RP is estimated by the event plane (EP), defined statistically in each event by the azimuth distribution of final-state particles in some acceptance. 
We now consider how to extract $v_m$ relative to a reaction plane estimated by an event plane in each event. Several different flow measures are implicitely defined in conventional flow analysis
\bea
v_m &\equiv& \overline{ \cos(m[\phi - \Psi_{r}])} ~~~\text{ideal case}, ~ 1/n \rightarrow 0 \\ \nonumber
\tilde v^2_m &\equiv& \langle \cos^2(m[\phi - \Psi_{r}])\rangle ~~~\text{unbiased estimate}  \\ \nonumber
\tilde q_m &\equiv& \langle \cos(m[\phi - \Psi_{m}])\rangle ~~~\text{self-pair bias} \\ \nonumber\tilde v'_m &\equiv& \langle \cos(m[\phi_i - \Psi_{mi}])\rangle ~~ \text {reduced bias}
\eea
$\tilde v'_m$ is the event-wise result of a conventional flow analysis. The ensemble average $\overline{\tilde v'_m}$ must be corrected for the  ``EP resolution'' which we now determine.


The basic event-wise quantities, starting with an integral over two-particle azimuth space,  are
\bea
\tilde V_m^2 &\equiv& \sum_{i \neq j}^{n,n-1} \vec u(m \phi_i) \cdot \vec u (m \phi_{j}) \\ \nonumber
&=& n(n-1) \langle \cos(m \phi_\Delta) \rangle \\ \nonumber
&=& n(n-1) \langle \cos^2(m [\phi - \Psi_r]) \rangle \\ \nonumber
&\equiv& n(n-1) \tilde v_m^2.
\eea
For the limiting case of subevents A and B with A a single particle the subevent azimuth vector complementary to particle $i$ is
\bea
\tilde {\vec Q}_{mi} &\equiv& \sum_{j \neq i}^{n-1} \vec u(m\phi_j) = \tilde Q_{mi} \vec u(m\Psi_{mi}).
 \eea
We make the following rearrangment
\bea
 \tilde V_m^2 &=& \sum_i^n \vec u(m\phi_i) \cdot \sum_{j \neq i}^{n-1} \vec u (m \phi_{j}) \\ \nonumber
&=& \sum_i^n \tilde Q_{mi} \cos(m[\phi_i - \Psi_{mi}]) \\ \nonumber 
 &\equiv&  n  \tilde Q'_{m} \langle \cos(m[(\phi -  \Psi'_m])\rangle \\ \nonumber
 n(n-1) \tilde v_m^2 &=& n \tilde Q'_m \, \tilde v'_m \\ \nonumber
\tilde v_m &=& \tilde v'_m\, \sqrt{\frac{n}{n-1}}\cdot \frac{\tilde Q'_m}{\tilde V_m  },
\eea
and, since $\tilde Q'_m \simeq \sqrt{n-1 + (n-1)(n-2) \langle \cos(m \phi_\Delta) \rangle}$,
we identify the EP resolution as
\bea
\langle \cos(m[\Psi'_m - \Psi_r]) \rangle &=& \sqrt{\frac{n-1}{n}}\cdot \frac{\tilde V_m}{\tilde Q'_m} \\ \nonumber
 &=& \sqrt{\frac{n-1}{n-2}}\cdot \frac{\tilde V'_m}{\tilde Q'_m},
\eea
where the primes refer to a subevent with $n-1$ particles. That expression, for full events with multiplicity $n$, is plotted in Fig.~\ref{fig1z} (right panel) for several choices of $n$. An $n$-independent universal curve on $V_m^2/\bar n$ is multiplied by $n$-dependent factor $\sqrt{n/(n-1)}$, where $n$ is the number of samples in the event or subevent. The dashed curve is the parameterization from~\cite{poskvol}.

The right panel indicates that for large $n \tilde v_m^2$ single-particle reaction-plane estimates can provide a ``flow'' measurement with manageable bias. For small $n \tilde v_m^2$ the EP resolution averaged over many events is itself a ``flow'' measurement, even though the reaction plane is inaccessible in any one event. $\tilde V_m^2$ is determined from the same underlying two-particle correlations by other means---the only difference is how pairs are grouped across events. In App.~\ref{subev} the EP resolution is determined for the case of equal subevents A and B 

From this exercise we conclude that event-plane determination is irrelevant for the measurement of cylindrical multipole moments (``flows''). Following a sequence of analysis steps in the conventional approach based on determination of and correction for the EP estimate, the event plane cancels out of the flow measurement. What results from the conventional method is {\em approximations} to signal components of power-spectrum elements which can be determined directly in the form $V^2_m/  \bar n = \bar n\, v^2_m $ obtained with the autocorrelation method, which defines an unbiased version of $v_m$. Event-plane determination {\em can be} useful for study of other event-wise phenomena in relation to azimuth multipoles.

\section{2D (joint) autocorrelations} \label{joint}

We now return to the more general problem of angular correlations on $(\eta_1,\eta_2,\phi_1,\phi_2)$. We consider the analysis of azimuth correlations based on autocorrelations, power spectra and cylindrical multipoles without respect to an event plane in the context of the general Fourier transform algebra presented in Sec.~\ref{foorier}. We seek a comprehensive method which treats $\eta$ and $\phi$ equivalently.

In conventional flow analysis there are two concerns beyond the measurement of flow in a fixed angular acceptance: a) study flow phenomena in multiple narrow rapidity bins to characterize the overall ``three-dimensional event shape,'' analogous to the sphericity ellipsoid but admitting of more complex shapes over some rapidity interval, and b) remove nonflow contributions to flow measurements as a systematic error. Maintaining adequate bin multiplicities to avoid bias is strongly emphasized in connection with a). The contrast between such individual Fourier decompositions on single-particle azimuth in single-particle rapidity bins and a comprehensive analysis in terms of two-particle joint angular autocorrelations is the subject of this section

\subsection{Stationarity condition}

\begin{figure}[h]
\includegraphics[keepaspectratio,width=1.65in]{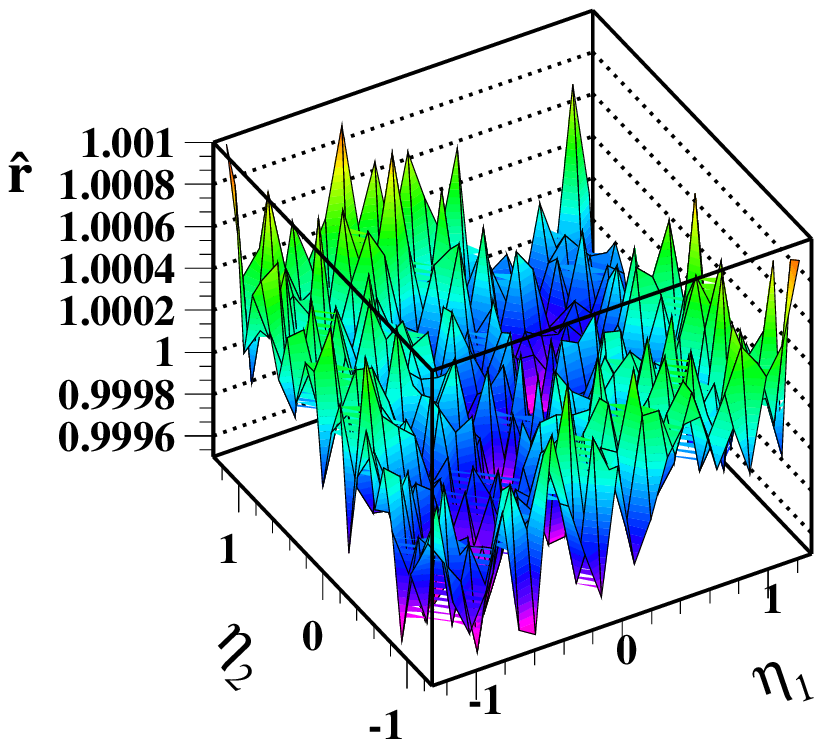}
\includegraphics[keepaspectratio,width=1.65in]{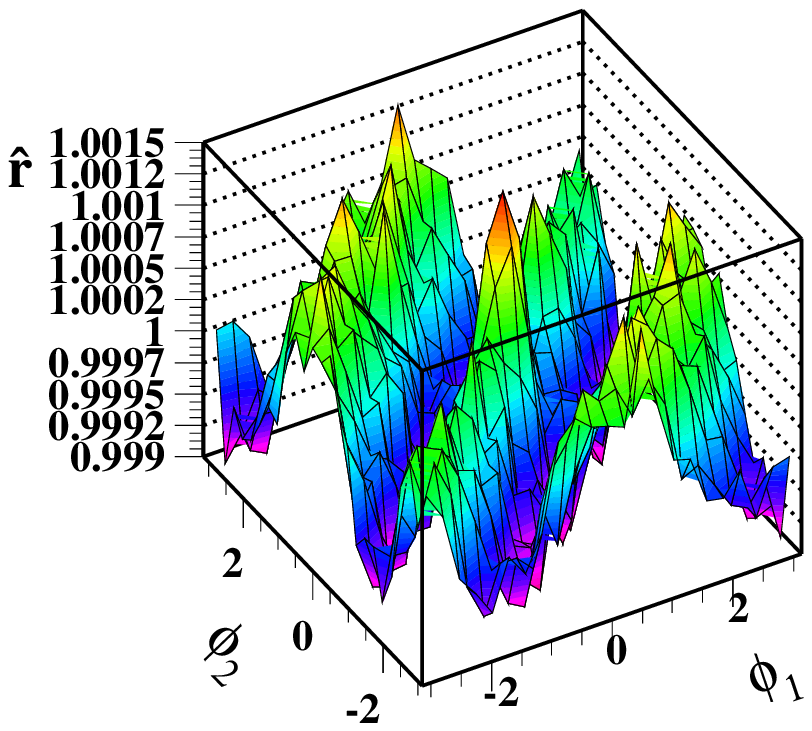}
\caption{\label{fig3}
Normalized like-sign pair-number ratios $\hat{r} = \rho/\rho_{ref}$ from central Au-Au collisions at 130 GeV for $(\eta_1,\eta_2)$ (left panel) and $(\phi_1,\phi_2)$ (right panel) showing {\em stationarity}---approximate invariance along sum diagonal $x_1 + x_2$. 
} 
\end{figure}

In Fig.~\ref{fig3} we show two-particle pair-density ratios $\hat r \equiv \rho / \rho_{ref}$ on $(\eta_1,\eta_2)$ (left panel) and $(\phi_1,\phi_2)$ (right panel) for mid-central 130 GeV Au-Au collisions~\cite{axialci}. The hat on $\hat r$ indicates that the number of mixed pairs in $\rho_{ref}$ has been normalized to the number of sibling pairs in $\rho$. In each case we observe approximate invariance along sum axes $\eta_\Sigma = \eta_1 + \eta_2$ and $\phi_\Sigma = \phi_1 + \phi_2$. In time-series analysis the equivalent invariance of correlation structure on the mean time is referred to as {\em stationarity}, implying that averaging pair densities along the sum axes loses no information. The resulting averages are autocorrelation distributions on difference axes $\eta_\Delta$ and $\phi_\Delta$.

In Fig.~\ref{fig3} (right panel) one can clearly see the $\cos(2 \phi_\Delta)$ structure conventionally associated with elliptic flow (quadrupole component). However, there are other contributions to the angular correlations which should be distinguished from multipole components, accomplished accurately by combining the two angular correlations into one joint angular autocorrelation. 

\subsection{Joint autocorrelation definition}

With the autocorrelation technique the dimensionality of pair density $\rho(\eta_1,\eta_2,\phi_1,\phi_2)$ can be reduced from 4D to 2D without information loss provided the distribution exhibits stationarity. Expressing  pair density $\rho(\eta_1,\eta_2,\phi_2,\phi_2) \rightarrow \rho(\eta_\Sigma,\eta_\Delta,\phi_\Sigma,\phi_\Delta)$ in differential form $d^4n/dx^4$ we define the joint autocorrelation on $(\eta_\Delta ,\phi_\Delta)$
\bea
\rho_A(\eta_\Delta ,\phi_\Delta) &\equiv& \frac {d^2}{d\eta_\Delta d\phi_\Delta} \left\langle \frac{d^2 n(\eta_\Sigma,\eta_\Delta ,\phi_\Sigma,\phi_\Delta)}{d\eta_\Sigma d\phi_\Sigma }\right\rangle_{\hspace{-.05in} \eta_\Sigma\,\phi_\Sigma }
\eea
by {\em averaging} the 4D density over $\eta_\Sigma$ and $\phi_\Sigma$ within a detector acceptance. The autocorrelation averaging on $\eta_\Sigma$ is equivalent to $\langle dn/d\eta \rangle_\eta \approx  n(\Delta \eta) / \Delta \eta$ at $\eta = 0$. The magnitude is still the 4D density $d^4n/d\eta_1 d\eta_2 d\phi_1 d\phi_2$, but it varies only on the two difference axes $(\eta_\Delta,\phi_\Delta)$.

\begin{figure}[h]
\includegraphics[keepaspectratio,width=1.65in]{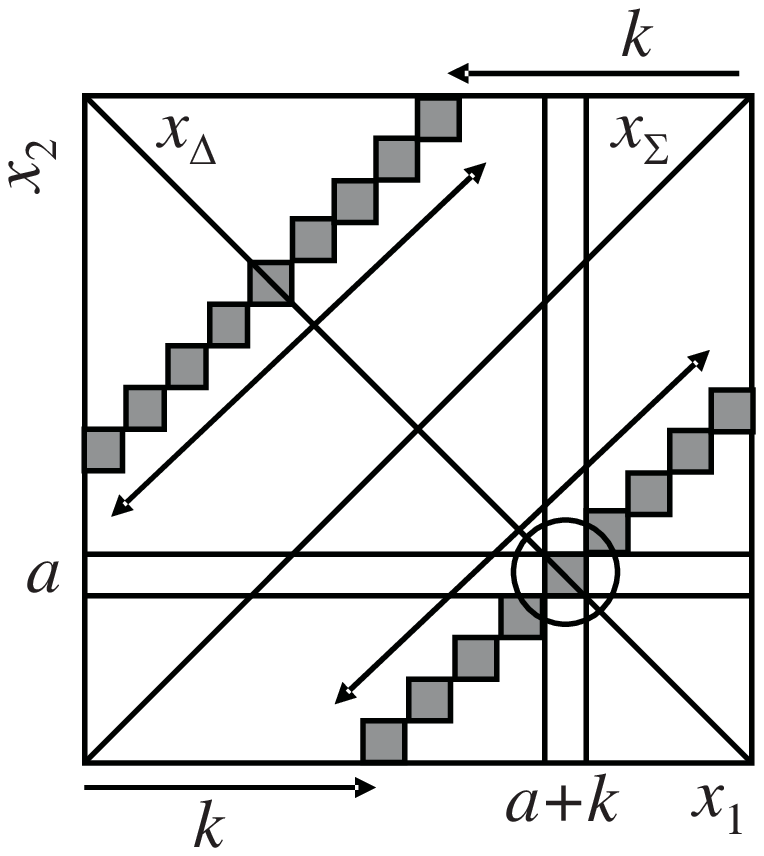}
\includegraphics[keepaspectratio,width=1.65in]{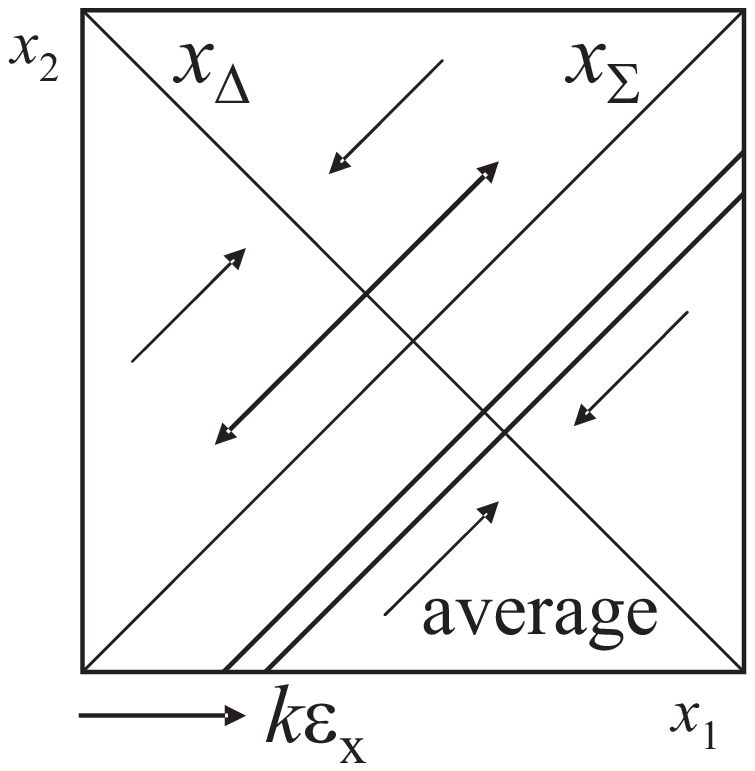}
\caption{\label{fig4}
Autocorrelation averaging schemes on $x_\Sigma = x_1 + x_2$ for a prebinned single-particle space (left panel) and for pairs accumulated directly into bins on the two-particle difference axis (right panel).
} 
\end{figure}

In Fig.~\ref{fig4} we illustrate two averaging schemes~\cite{inverse}. In the left panel we show the averaging procedure applied to histograms on $(x_1,x_2)$ as in Fig.~\ref{fig3}. Index $k$ denotes the position of an averaging diagonal on the difference axis. In the right panel we show a definition involving pair cuts applied on the difference axes. Pairs are histogrammed directly onto $(\eta_\Delta,\phi_\Delta)$. Periodicity on $\phi$ implies that the averaging interval on $\phi_\Sigma / 2$ is $2\pi$ independent of $\phi_\Delta$. However, $\eta$ is not periodic and the averaging interval on $\eta_\Sigma/2$ is $\Delta \eta - |\eta_\Delta |$, where $\Delta \eta$ is the single-particle $\eta$ acceptance~\cite{inverse}. The autocorrelation value for a given $x_\Delta$ is the bin sum along that diagonal divided by the averaging interval.

\subsection{Autocorrelation examples}

\begin{figure}[h]
\includegraphics[keepaspectratio,width=1.65in]{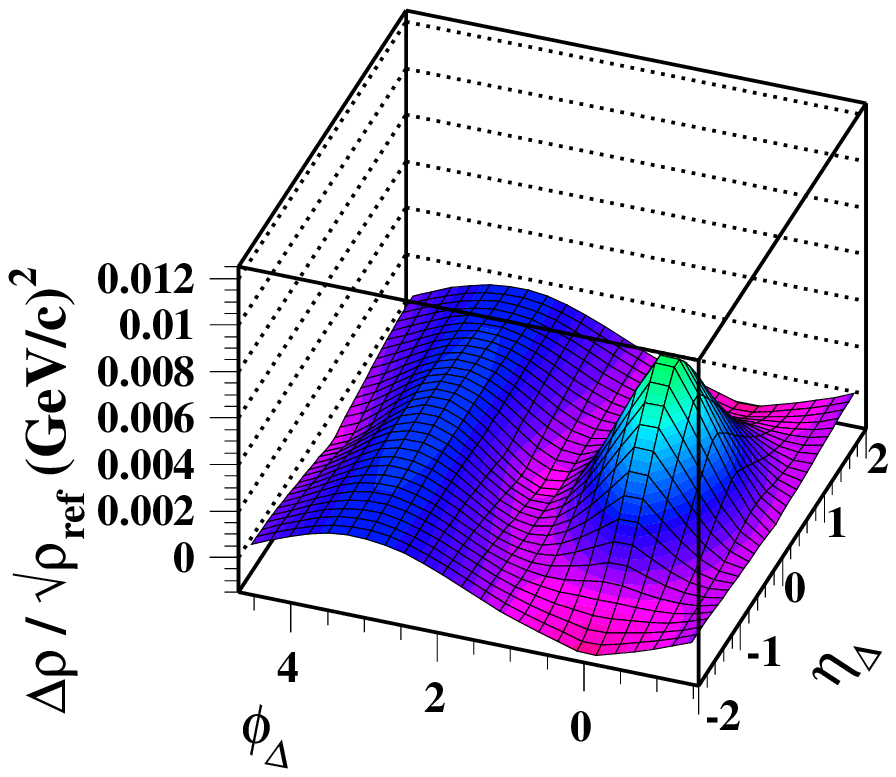}
\includegraphics[keepaspectratio,width=1.65in]{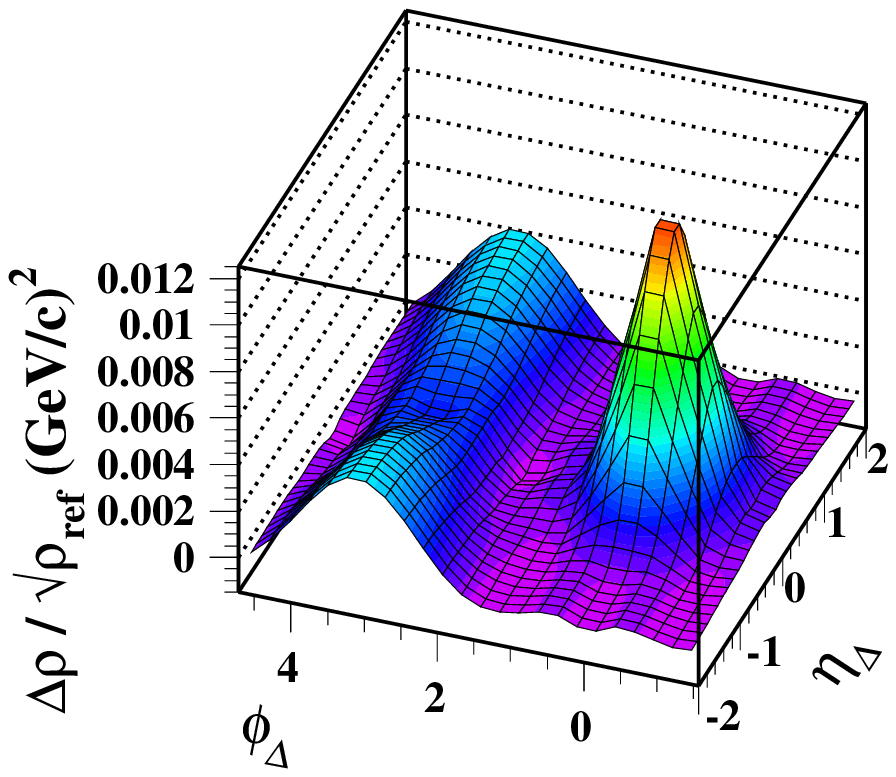}
\caption{\label{fig5}
2D $p_t$ angular autocorrelations from Hijing central Au-Au collisions at 200 GeV for quench on (left panel) and  quench off (right panel) simulations. 
} 
\end{figure}

In Fig.~\ref{fig5} we show $p_t$ joint angular autocorrelations for Hijing central Au-Au collisions at 200 GeV~\cite{hijscale}. The left panel shows quench-on collisions. The right panel shows quench-off collisions. Aside from an amplitude change Hijing shows little change in correlation structure from N-N to central Au-Au collsions. Those results can be contrasted with examples from an analysis of real RHIC data~\cite{hijscale} shown in Fig.~\ref{fig6a}.

In N-N or p-p collisions there is no {\em obvious} quadrupole component. However, that possibility is not quantitatively excluded by the data and requires careful fitting techniques. For correlation structure which is not sinusoidal there is no point to invoking the Wiener-Khintchine theorem on $\phi_\Delta$ and transforming to a power spectrum. Instead, model functions specifically suited to such structure (e.g., 1D and 2D gaussians) are more appropriate. The power-spectrum model is appropriate for {\em some} parts of $\rho_A$, depending on the collision system. We consider hybrid decompositions in Sec.~\ref{autostruct}.

\subsection{Comparison with conventional methods}

We can compare the power of the autocorrelation technique with the conventional approach based on EP estimation in single-particle $\eta$ bins. There are concerns in the single-particle approach about bias or distortion from low bin multiplicities. In contrast, the autocorrelation method is applicable to any collision system with any multiplicity, as long as {\em some} pairs appear in {\em some} events. The method is minimally biased, and the statistical error in $\Delta \rho_A / \sqrt{\rho_{ref}}$ is determined only by the number of bins and the total number of events in an ensemble.

There is interest within the conventional context in ``correlating'' flow results in different $\eta$ bins. ``A three-dimensional event shape can be obtained by correlating and combining the Fourier coefficients in different longitudinal windows"~\cite{volzhang}.  That goal is automatically achieved with the joint autocorrelation technique but has not been implemented with the conventional method. The content of an autocorrelation bin at some $\eta_\Delta$ represents a covariance averaged over all bin pairs at $\eta_a$ and $\eta_b$ satisfying $\eta_\Delta = \eta_a - \eta_b$. If a flow component is isolated (e.g., $V_2^2 / \bar n$), a given autocorrelation element represents normalized covariance $\overline{n_a\, n_b \tilde v^2_{2 a b}} / \sqrt{\bar n_a \bar n_b}$, where $\tilde v_{mab}^2$ would be the event-wise result of a `subevent' or `scalar-product' analysis between bins $a$ and $b$. Averaged over many events one obtains good resolution on rapidity and azimuth for arbitrary structures, without model dependence or bias.

\section{Flow and minijets (nonflow)} \label{nonflow}

The relation between azimuth multipoles and minijets is a critically important issue in heavy ion physics which deserves precise study. We should carefully compare multipole structures conventionally attributed to hydrodynamic flows and parton fragmentation dominated by minijets in the same analysis context. The best arena for that comparison is the 2D (joint) angular autocorrelation and corresponding power-spectrum elements. Before proceeding to autocorrelation structure we consider nonflow in the conventional flow context

\subsection{Nonflow and conventional flow analysis}

In conventional flow analysis azimuth correlation structure is simply divided into `flow' and `nonflow,' where the latter is conceived of as non-sinusoidal structure of indeterminant origin. The premise is that all sinusoidal structure represents flows of hydrodynamic origin. It is speculated that nonflow is due to resonances, HBT and jets, including minijets. Various properties are assigned to nonflow which are said to distinguish it from flow~\cite{2004}. Nonflow is by definition non-sinusoidal and is not correlated with the RP, thus it can appear perpendicular to the RP. 

The multiplicity dependence of nonflow is said to be quite different from flow, where ``multiplicity dependence'' can sometimes be read as centrality dependence. For instance,  $\tilde {\vec Q}_{ma} \cdot \tilde {\vec Q}_{mb} = \tilde V_{ma}\, \tilde V_{mb} \cos(\Psi_{ma} - \Psi_{mb})$ if A, B are disjoint, since there are no self pairs. The ensemble average then measures covariance $V_{mab}^2$. For $m  =2$ it is claimed that the nonflow component of $V_{2ab}^2$ is $  \propto \bar n \, c$~\cite{poskvol}. Therefore, $V_{2}^2 / \bar n \propto c$, a constant for nonflow---no centrality dependence. But  $V_{m}^2 / \bar n \propto \Delta \rho_A / \sqrt{\rho_{ref}}$ which, for the minijets dominating nonflow, is very strongly dependent on centrality~\cite{axialci,ptscale}; the conventional assumption is incorrect. The above notation is inadequate because minijets (nonflow) should not be included in the Fourier power spectrum. They should be modeled by different functional forms which we consider in the next section.

\subsection{Cumulants}

Another strategy for isolating flow from nonflow is to use higher cumulants~\cite{borg}. The basic assumption (a physical correlation model) is that flow sinusoids are collective phenomenon characteristic of almost all particles, whereas nonflow is a property only of pairs, termed ``clusters.'' That scenario is said to imply that $v_m$ should be the same no matter what the multiplicity, whereas nonflow should fall off as some inverse power of $n$.
    
For instance, by subtracting $v_2$[4] (four-particle cumulant) from $v_2$[2] (two-particle cumulant) one should obtain ``nonflow'' as the difference ({\em cf.} Eq. (10) of ~\cite{2004}). In Fig.~31 of~\cite{2004} we find a plot of $g_2 = N_{part} \, (v^2_2 [2] - v^2_2 [4]) \propto N_{part} / n_{ch} \times \Delta \rho / \sqrt{\rho_{ref}} $. Multiplying $g_2$ by $n_{ch}^2 / N_{part}^2$ we obtain a measure of minijet correlations per participant pair. That `nonflow' component increases rapidly with centrality (and therefore $n$), consistent with actual measurements of minijet centrality trends. The incorrect factor in the definition of $g_2$ removes a factor 2x increase from peripheral to central in the minijet centrality trend, thus suppressing the centrality dependence of `nonflow.' 

\subsection{Counterarguments}

The conventional flow analysis method requires a complex strategy to distinguish flow from nonflow in projections onto 1D azimuth difference $\phi_\Delta$. An intricate and fragile system results, with multiple constraints and assumptions. The assumptions are not {\em a priori} justified, and must be tested. `Flow' isolated with those assumptions can and does contain substantial systematic errors.

Claims about the multiplicity (centrality) dependence of `nonflow' (independent of centrality or slowly varying) are {\em unsupported speculations without basis in  experiment}. In fact, detailed measurements of minijet centrality dependence~\cite{axialci,ptscale} are quite inconsistent with typical assumptions about nonflow. Multiplicity (centrality) dependence of flow measurements is further compromised by biases resulting from improper statistical methods, especially true for small multiplicities or peripheral collisions. Such biases can masquerade as physical phenomena.

Finally, it is assumed that nonflow has no correlation with the RP, thus implying the ability of and need for the EP to distinguish flow from nonflow. But nonflow (minijets) {\em should} be strongly correlated with the EP (jet quenching), and such correlations should be measured. That is the main subject of paper II in this two-part series. Precise decomposition of angular correlations into `flow' sinusoids and minijet structure is  realized with 2D joint angular autocorrelations combined with proper statistical techniques.

Non-flow is a suite of physical phenomena, each worthy of detailed study. In the conventional approach this physics is seen in limited ways by various projections and poorly-designed measures and described mainly by speculation. With more powerful analysis methods it is possibl to separate flow from the various sources of `nonflow' reliably and identify those sources as interesting physical phenomena.

\section{Structure of the joint angular autocorrelation  in A-A collisions} \label{autostruct}

We now return to the 2D angular autocorrelation. By separating its structure into a few well-defined components we obtain an accurate separation of multipoles, minijets and other phenomena. Minijets and ``flows'' can be compared quantitatively within the same analysis context. 

Each bin of an autocorrelation is a comparison of two ``subevents.'' The notional term ``subevent'' represents a partition element in conventional math terminology (e.g., topology, {\em cf.} Borel measure theory). An ``event'' is a distribution in a bounded region of momentum space (detector acceptance), and a subevent is a partition element thereof. A distribution can be partitioned in many ways: by random selection, by binning the momentum space, by particle type, etc. A uniform partition is a binning, and the set of bin entries is a histogram.  

A bin in an angular autocorrelation represents an average over all bin pairs in single-particle space separated by certain angular differences $(\eta_\Delta,\phi_\Delta)$. The bin contents represent normalized covariances averaged over all such pairs of bins. The notional ``scalar product method,'' relating two subevents in conventional flow analysis, is already incorporated in conventional mathematical methods developed over the past century as covariances in bins of an angular autocorrelation. Using 2D angular autocorrelations we easily and accurately separate nonflow from flow. ``Nonflow''  so isolated has revealed the physics of minijets---hadron fragments from the low-momentum partons which dominate RHIC collisions.

\subsection{Minijet angular correlations}

Minijet correlations are  equal partners with multipole correlations on difference-variable space $(\eta_\Delta,\phi_\Delta)$. Minimum-bias jet angular correlations (dominated by minijets) have been studied extensively for p-p and Au-Au collisions at 130 and 200 GeV~\cite{ppcorr,axialci,ptscale,edep}. Those structures dominate the ``nonflow'' of conventional flow analysis. In p-p collisions minijet structure---a same-side peak (jet cone) and away-side ridge uniform on $\eta_\Delta$---are evident for hadron pairs down to 0.35 GeV/c for each hadron. Parton fragmentation down to such low hadron momenta is fully consistent with fragmentation studies over a broad range of parton energies (e.g., LEP, HERA)~\cite{fragfunc}.

\begin{figure}[h]
\includegraphics[keepaspectratio,width=1.65in]{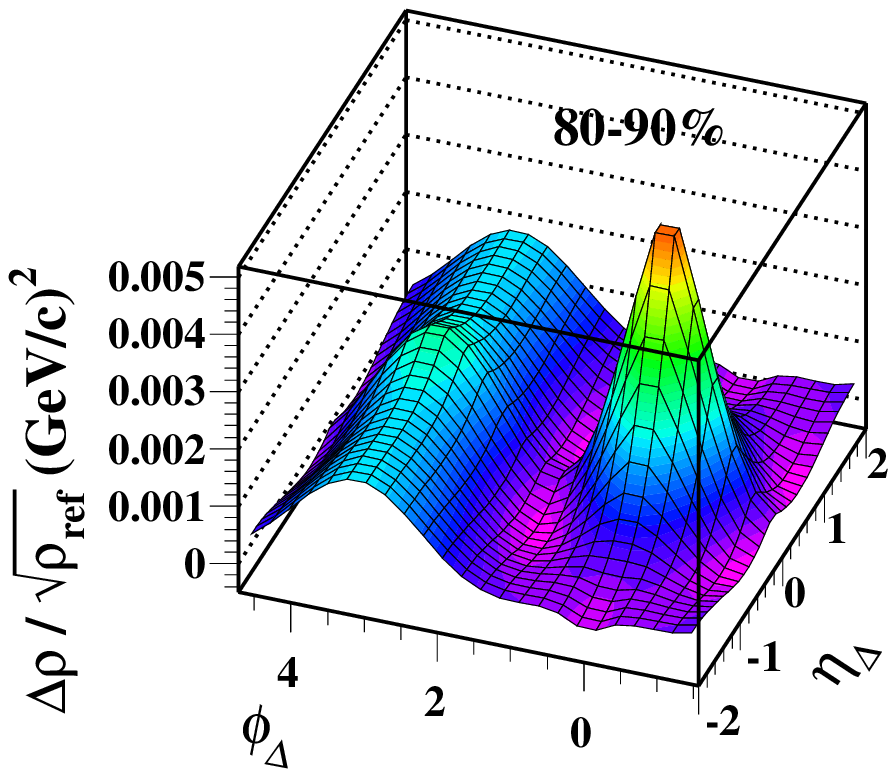}
\includegraphics[keepaspectratio,width=1.65in]{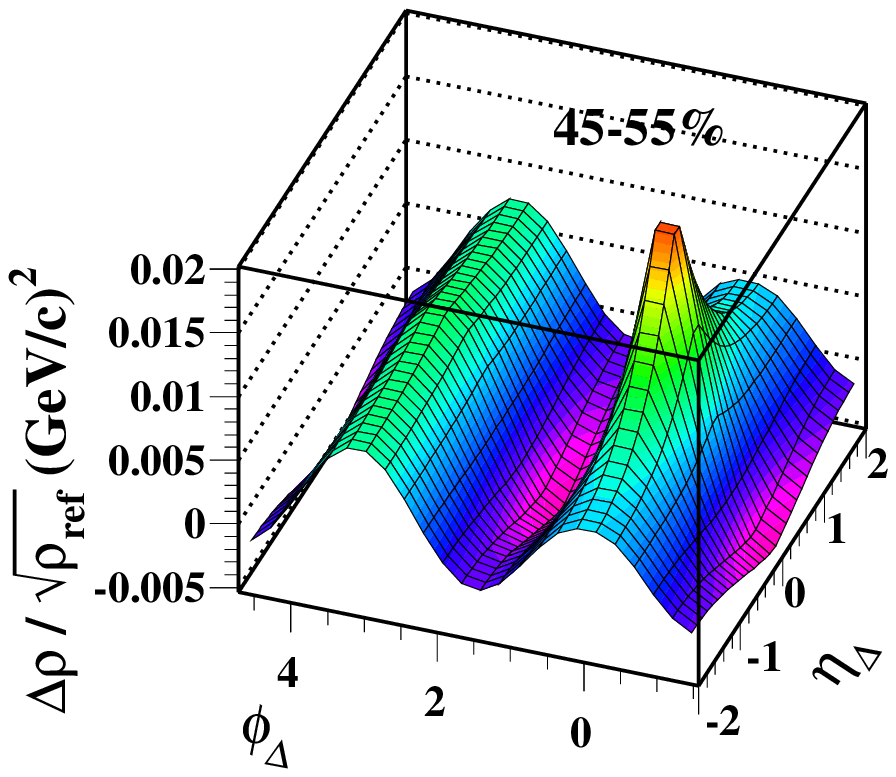}
\includegraphics[keepaspectratio,width=1.65in]{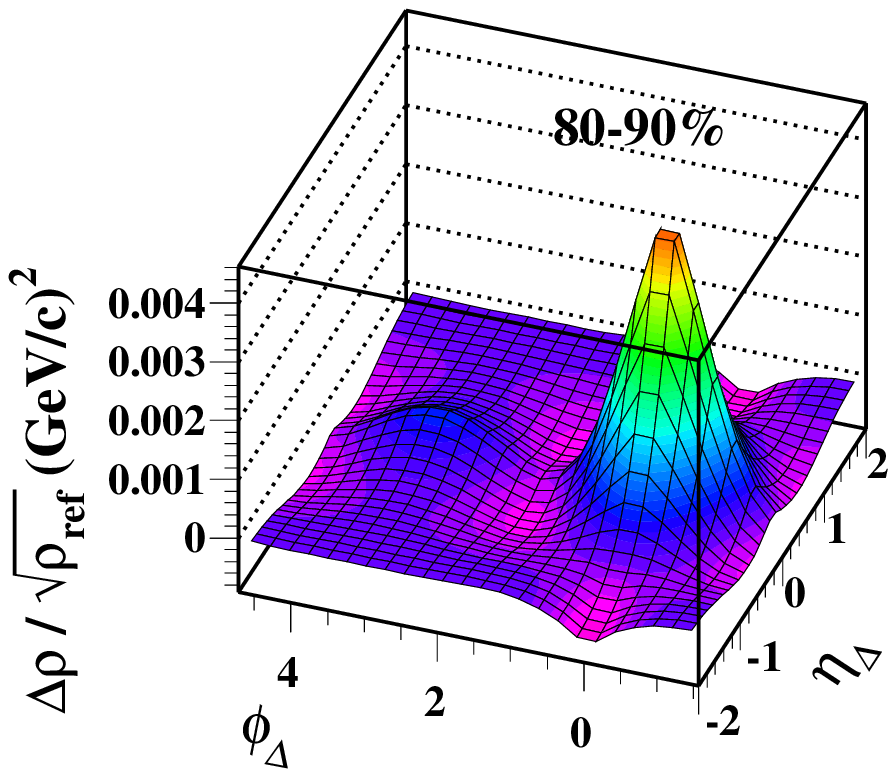}
\includegraphics[keepaspectratio,width=1.65in]{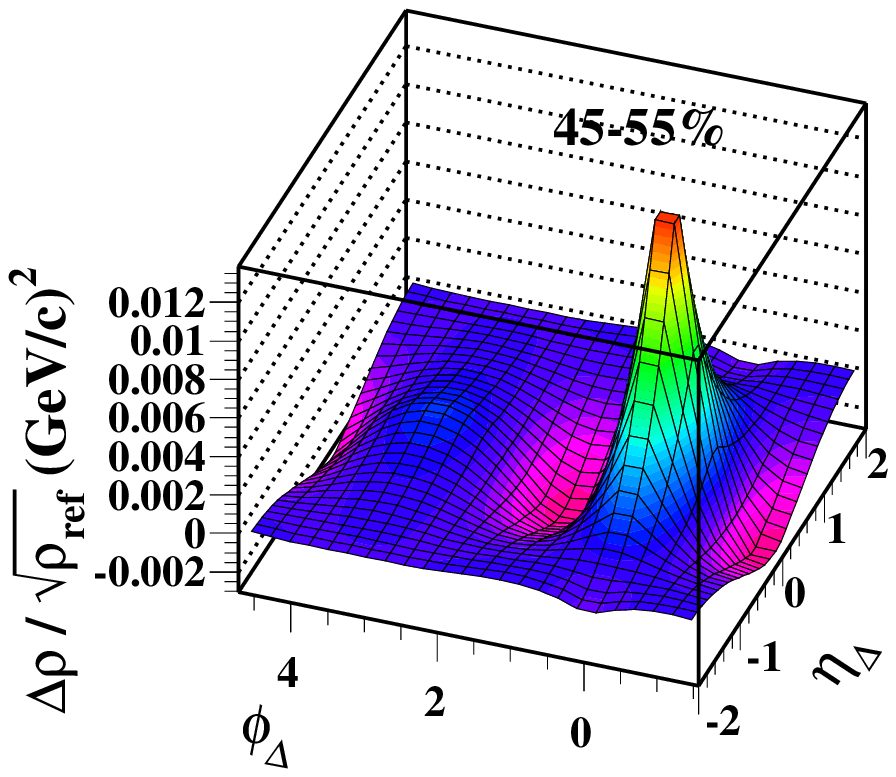}
\caption{\label{fig6a}
2D $p_t$ angular autocorrelations from Au-Au collisions at 200 GeV for 80-90\% central collisions (left panels) and 45-55\% central collisions (right panels). In the lower panels sinusoids $\cos(\phi_\Delta)$ and $\cos(2\phi_\Delta)$ have been subtracted to reveal ``nonflow'' structure. 
}  
\end{figure}

In Fig.~\ref{fig6a} we show autocorrelations obtained by inversion of $p_t$ fluctuation scale (bin-size) dependence~\cite{ptscale}. The upper-left panel is 80-90\% central and the upper-right panel is 45-55\% central Au-Au collisions. Correlation structure is dominated by a same-side peak and multipole structures (sinusoids). Subtracting the sinusoids reveals the minijet structure in the bottom panels and illustrates the precision with which flow and nonflow can be distinguished. The negative structure surrounding the same-side peak at lower right is an interesting and unanticipated new feature~\cite{ptscale}.

\subsection{Decomposing 2D angular autocorrelations: a controlled comparison}

Based on extensive analysis~\cite{ppcorr,hijscale,ptscale,edep} we find three main contributions to angular correlations in RHIC nuclear collisions: 1) transverse fragmentation (mainly minijets), 2) longitudinal fragmentation (modeled as ``string'' fragmentation), 3) azimuth multipoles (flows). Longitudinal fragmentation plays a reduced role in heavy ion collisions. In this study we focus on the interplay between 1) and 3), transverse parton fragmentation and azimuth multipoles, as the critical analysis issue for azimuth correlations in A-A collisions. 

The 2D joint autocorrelation $\rho_A(\eta_\Delta,\phi_\Delta)$ is the basis for decomposition. The criteria for distinguishing azimuth multipoles from minijet structure are $\eta_\Delta$ dependence and sinusoidal $\phi_\Delta$ dependence. Structure with sinusoidal $\phi_\Delta$ dependence and $\eta_\Delta$ invariance is assigned to azimuth multipoles. Other structure, varying generally on $(\eta_\Delta,\phi_\Delta)$, is assigned {\em in this exercise} to minijets.  We adopt the decomposition
\bea
\rho_A(\eta_\Delta,\phi_\Delta) = \rho_j(\eta_\Delta,\phi_\Delta) + \rho_m(\phi_\Delta),
\eea 
where $j$ represents (mini)jets and $m$ represents multipoles. That decomposition is reasonable within a limited pseudorapidity acceptance, e.g., the STAR TPC acceptance~\cite{startpc}. Over a larger acceptance other separation criteria must be added.

\begin{figure}[b]
\includegraphics[keepaspectratio,width=3.3in]{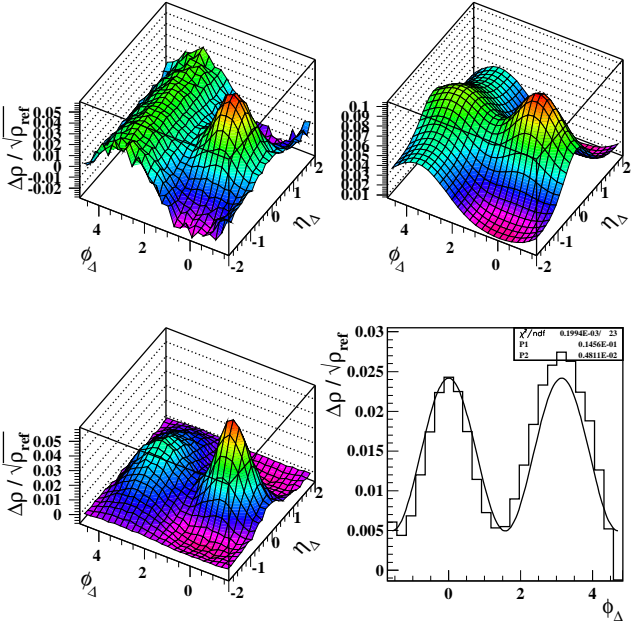}
\caption{\label{fig6}
Simulated three-component 2D angular autocorrelation for 80-90\% central Au-Au collisions at 200 GeV (upper left), model of data distribution from fitting (upper-right), autocorrelation with $eta$-acceptance triangle imposed (lower left) and sinusoid fit to 1D projection of lower-left panel (lower-right).
}  
\end{figure}

To illustrate the separation process we construct artificial autocorrelations combining flow sinusoids and minijet structure with centrality dependence taken from measurements. We add statistical noise appropriate to a typical event ensemble of a few million  Au-Au collisions. We then fit the autocorrelations with model functions and $\chi^2$ minimization. We compare the resulting fit parameters with the input parameters. We then project the 2D autocorrelations onto $\phi_\Delta$ and fit the results with a sinusoid. The result of the 1D fit represents the product of a conventional flow analysis if the EP resolution is perfectly corrected and there is no statistical bias in the method. We then compare the resulting sinusoid amplitudes.

In Fig.~\ref{fig6} we show an analysis for 80-90\% central Au-Au collisions, which are nearly N-N collisions. The upper-right panel is an accurate model of data from p-p collisions~\cite{ppcorr}. The upper-left panel is the simpler representation for this exercise with added statistical noise. The difference in constant offsets is not relevant to the exercise. At lower left is the distribution ``seen'' by a conventional flow analysis, which simply integrates over (projects) the $\eta$ dependence. The projection includes a triangular acceptance factor on $\eta_\Delta$ imposed on the joint autocorrelation, resulting in distortions that affect the sinusoid fit. The lower-right panel is the projection onto $\phi_\Delta$ with $\chi^2$ fit. The 1D fit gives $\Delta \rho[2]/\sqrt{\rho_{ref}} = 0.0052$, compared to 0.0013 from the 2D fit

\begin{figure}[b]
\includegraphics[keepaspectratio,width=3.3in]{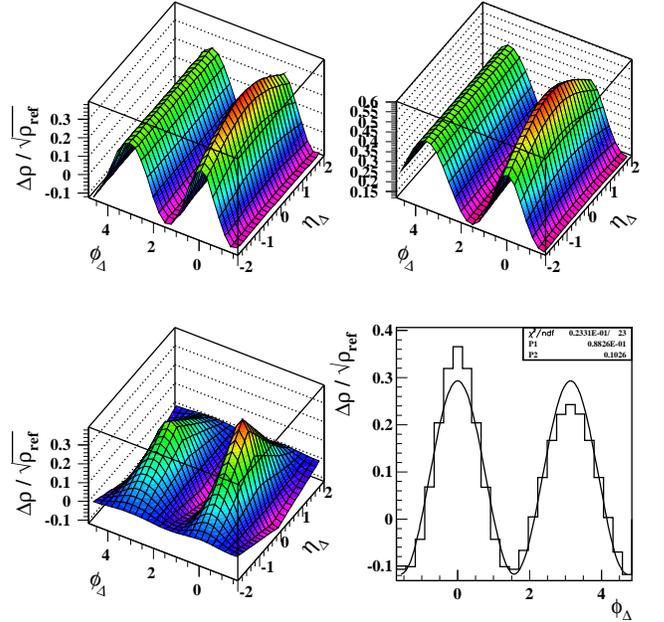}
\caption{\label{fig7}
Same as the previous figure but for 5-10\% central Au-Au collisions at 200 GeV. Note the pronounced effect of the $\eta$-acceptance triangle in the lower-left panel (relative to the upper-right panel) resulting from projection of space $(\eta_1,\eta_2)$ onto its difference axis.
}  
\end{figure}

In Fig.~\ref{fig7} we show the same analysis applied to 5-10\% Au-Au collisions. The model distribution, derived from observed data trends, is dominated by a dipole term $\propto \cos(\phi_\Delta)$ and an elongated same-side jet peak, although the {\em combination} closely mimics a quadrupole $\propto \cos(2\phi_\Delta)$ (elliptic flow). The lower-left panel shows the effect of the $\eta$-acceptance triangle on $\eta_\Delta$ applied to the upper-right panel which is implicit in any projection onto $\phi_\Delta$ (obvious in this 2D plot). The resulting projection on $\phi_\Delta$ is shown at lower right. The 1D fit gives $\Delta \rho[2]/\sqrt{\rho_{ref}} = 0.103$, compared to 0.060 from the 2D fit. The differences between 1D and 2D fits are much larger than the differences between input and fitted parameters in the previous exercise. They reveal the limitations of conventional flow analysis.

\begin{figure}[h]
\includegraphics[keepaspectratio,width=3.3in]{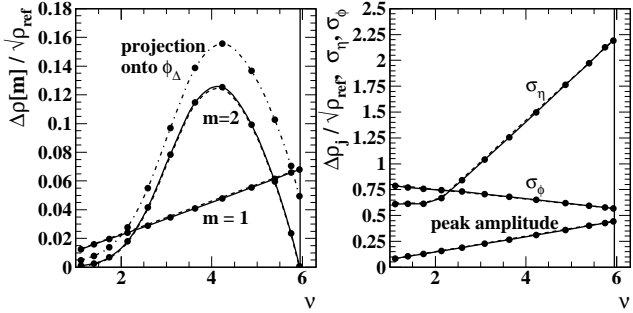}
\caption{\label{fig8} A parameter summary for the previous two figures. The solid curves are input model parameters for $m = 1$ and $m = 2$ sinusoids $\cos(m\phi_\Delta)$ and a same-side 2D gaussian with two widths and peak amplitude which approximate 200 GeV Au-Au collisions. The dashed curves are the results of fits to the model 2D autocorrelations exhibiting excellent accuracy. The dash-dot curve represents 1D fits to projections on $\phi_\Delta$ (previous lower-right panels) corresponding to conventional flow analysis.
}  
\end{figure}

In Fig.~\ref{fig8} we give a summary of results for twelve centrality classes, nine 10\% bins with mean values 95\% $\cdots$ 15\%, plus two 5\% bins with mean values 7.5\% and 2.5\%. The twelfth class is $b = 0$ (0\%), constructed by extrapolating the parameterizations from data. The solid curves and points represent the parameters inferred from 2D fits to the joint angular autocorrelations. The dashed curves represent the model parameters used to construct the simulated distributions. There is good agreement at the percent level. 

The fits to 1D projections on $\phi_\Delta$ (dash-dot curve) however differ markedly from the 2D fit results and the input parameters. The differences are very similar to the changes of conventional flow measurements with different strategies to eliminate ``nonflow.'' This exercise demonstrates that with the 2D autocorrelation there is no guesswork. We can distinguish the multipole contributions from the minijet contributions. The 2D angular autocorrelation provides precise control of the separation.

\subsection{$v_2$ in various contexts} \label{various}

\begin{figure}[h]
\includegraphics[keepaspectratio,width=3.3in]{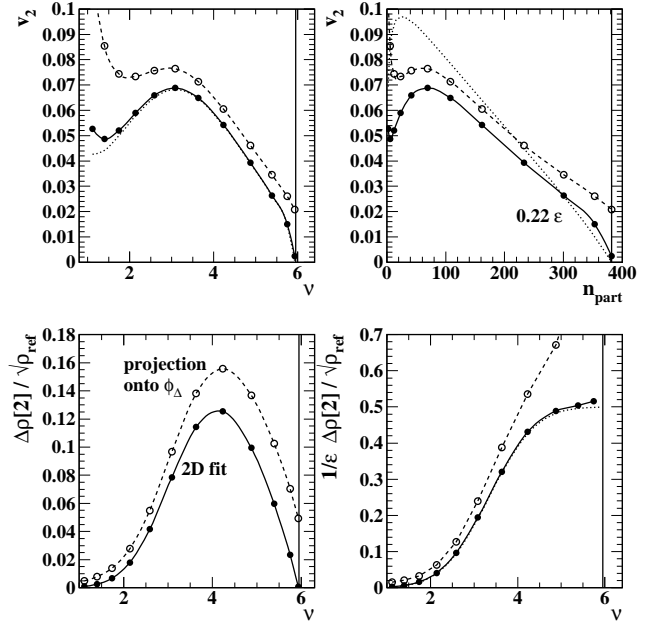}
\caption{\label{fig9}
The quadrupole component in various plotting contexts derived from the previous model exercise. Conventional flow measure $v_2$ is shown in the upper panels. $\Delta \rho[2] / \sqrt{\rho_{ref}}$ based on Pearson's normalized covariance is shown in the lower panels. The upper-right panel shows $v_2$ {\em vs} $n_{part}$ in the conventional plotting format. The lower-left panel repeats the left panel of the previous figure, with conventional 1D fit results shown as the dashed curve. Dashed and solid curves correspond in all panels. $\nu$ estimates the mean N-N encounters per participant pair. The dotted curve at lower right is the error function used to generate the model quadrupole amplitudes. It's correspondent for $v_2$ is at upper left. The dotted curve at upper right is eccentricity $\epsilon$ from a parameterization.
}   
\end{figure}

In Fig.~\ref{fig9} we contrast the results of 1D conventional flow analysis (dashed curves) and extraction of the quadrupole amplitude from the 2D angular autocorrelation (solid curves). We make the correspondence
\bea
\frac{\Delta \rho_A[2] }{ \sqrt{\rho_{ref}}} = \frac{V_2^2}{2 \pi \,\bar n} = \frac{\bar n v_2^2}{2 \pi}
\eea 
 among variables, with $\bar n / 2\pi$ modeling $\rho_0 = \overline{d^2n/ d\eta d\phi}$. That expression defines a minimally-biased $v_2$. We make the comparison in four plotting formats. The upper-right panel is most familiar from conventional flow analysis, where $v_2$ is plotted {\em vs} participant nucleon number. The trends can be compared with Fig. 13 of~\cite{2002} (open circles {\em vs} solid stars). Also included in that panel is the trend $v_2 \sim 0.22\, \epsilon$ predicted by hydro for thermalized A-A collisions at 200 GeV (dotted curve,~\cite{jceps}).

The remaining panels are plotted on parameter $\nu = 2 n_{bin} / n_{part}$, the ratio of N-N binary {\em encounters} to participant pairs which estimates the mean participant path length in number of encountered nucleons, a geometrical measure. Comparing the upper panels we see that $\nu$ treats peripheral and central collisions equitably, whereas $n_{part}$ or $n_{charge}$ compresses the important peripheral region into a small interval.

In the lower-left panel we plot per-particle density ratio $\Delta \rho_A[2] / \sqrt{\rho_{ref}}$ {\em vs} $\nu$. That quantity, when extracted from 2D fits, rises from near zero for peripheral collisions to a maximum for mid-central collisions, falling toward zero again for $b = 0$. In contrast to $v_2$, which is the square root of a per-pair correlation measure, the per-particle density ratio reflects the trend for ``flow'' in the sense of a current density. ``Flow'' is small for peripheral collisions and grows rapidly with increasing nucleon path length. The trend with centrality is intuitive. The values obtained from the 1D projection per conventional flow analysis (dashed curve) are consistently high, especially for central collisions, exhibiting a strong systematic bias.
The 1D fit procedure (identical to the ``standard'' and two-particle flow methods) confuses minijet structure with quadrupole structure. 

In the lower-right panel we show the density ratio divided by initial spatial eccentricity $\epsilon$ defined by a parameterization derived from a Glauber simulation and plotted as the dotted curve in the upper-right panel~\cite{jceps}. 
The trend from 2D fits (solid curve) is closely approximated by a simple error function (dotted curve) with half-maximum point at the center of the $\nu$ range. In fact, the dotted curve is the basis for generating the input quadrupole amplitudes for our model, and the small deviation of the solid curve from the dotted curves in upper-left and lower-right panels reveals the systematic error or bias in the 2D fitting procedure ($\sim 20$\% at $\nu \sim 1$, $< 5$\% at $\nu = 5.8$). The dashed curves from the conventional 1D fits show large relative deviations from the input trend, especially for peripheral collisions where the emergence of collectivity is of interest and for central collisions where the issue of thermalization is most important. 

Comparing the solid curve to existing $v_2$ data in the format of the upper-right panel shown in~\cite{2002} (Fig.~13) indicates that our simple formulation at lower right (dotted curve) is roughly consistent with analysis of flow data based on four-particle cumulants. The result from data generated with the same model and analyzed with the conventional flow analysis method (dashed curve in upper-right panel) also agrees with the conventional method applied to real RHIC data. Our model  may therefore indicate an underlying simplicity to the quadrupole mechanism which is not hydrodynamic in origin. The model centrality trend for $v_2$ is certainly {\em inconsistent} with the hydro expectation $v_2 \propto \epsilon$~\cite{ollitrault,volposk}, as demonstrated in the upper-right panel ({\em cf.} App.~\ref{centro}).

\section{Discussion}

\subsection{Conventional flow analysis}

The overarching premise of conventional flow analysis is that in the azimuth distribution of each collision event lies evidence of collective phenomena which must be discovered to establish event-wise thermalization. The $1/n \rightarrow 0$ hydro limit shapes the analysis strategy, and flow manifestations are the principal goal. Finite event multiplicities are seen as a major source of systematic error, as are correlation structures other than flow. Multiple strategies are constructed to deal with non-flow and finite multiplicities. A stated advantage of the conventional method is that the Fourier coefficients {\em can} be corrected. The great disadvantage is they {\em must} be corrected.

From the perspective of two-particle correlation analysis, especially in the context of autocorrelations and power spectra, the conventional program leaves much to be desired. The conventional analysis is essentially an attempt to measure two-particle correlations with single-particle methods combined with RP estimation, similar to the use of trigger particles in high-$p_t$ jet analysis. But, by analysis of the algebraic structure of conventional flow analysis we have demonstrated that RP estimation does not matter to the end result. Without a proper statistical reference conventional analysis results contain extraneous contributions from the statistical reference which are partially `corrected' in a number of ways. The improper treatment of random variables incorporates sources of multiplicity-dependent bias in measurements, and the final results are questionable.

Flow measure $v_m$ is nominally the square root of {\em per-pair} correlation measure $V^2_m / \overline{n(n-1)}$. $v_m$ centrality trends are thus nonintuitive and misleading (e.g., ``elliptic flow'' decreases with increasing A-A centrality). The situation is similar to {\em per-pair} fluctuation measure $\Sigma_{p_t}$, which provides a dramatically misleading picture of $p_t$ fluctuation dependence on collision energy~\cite{edep}. In contrast, {\em per-particle} correlation measures provide intuitively clear results and often make dynamical correlation mechanisms immediately obvious. In particular, the mechanisms behind ``nonflow'' in the form of minijets are clearly apparent when correlations are measured by per-particle normalized covariance density $\Delta \rho / \sqrt{\rho_{ref}}$ in a 2D autocorrelation. 

\subsection{Autocorrelations and nonflow}

In conventional flow analysis it is proposed to measure flow in narrow rapidity bins (``strips'') so as to develop a three-dimensional picture of event structure. However, there has been little implementation of that proposal. In contrast, the joint angular autocorrelation {\em by construction} contains all possible covariances among pseudorapidity bins within a detector acceptance. The ideal of full event characterization is thereby realized.

The angular autocorrelation is the optimum solution to a geometry problem---how to reduce the six-dimensional two-particle momentum space to two-dimensional subspaces with minimum distortion or information loss. The autocorrelation is the {\em unique solution} to that problem, involving no model assumptions. The ensemble-averaged angular autocorrelation contains all the correlation information obtainable from a conventional flow analysis, but with negligible bias and no sensitivity to individual event multiplicities.

Because it is a two-dimensional representation the angular autocorrelation is far superior for separating ``flow'' (multipoles) from ``nonflow'' (minijets), as we have demonstrated. A simple exercise demonstrates that separation is complete at the percent level, whereas the conventional method admits crosstalk at the tens of percent level. Precise separation leads to new physics insights from the multipoles and minijets so revealed.

\subsection{Collision centrality dependence}

Collision centrality dependence is of critical importance in the comparison of flow and minijets. Parton collisions and hydrodynamic response to early pressure have very different dependence on impact parameter and collision geometry, especially for peripheral collisions. Peripheral A-A collisions should approach p-p (N-N) collisions, and correlation structure may change rapidly in mid-peripheral collisions as collective phenomena develop there. The possible onset of collective behavior in mid-peripheral collisions and reduction in more central collisions are of major importance for understanding the relation of minijets to flow. The conventional flow analysis method is severely limited for peripheral collisions. In contrast, correlation measure $\Delta \rho / \sqrt{\rho_{ref}}$, centrality measure $\nu$ and associated centrality techniques described in~\cite{centmeth} are uniquely adapted to cover all centrality regions down to N-N with excellent accuracy.

\subsection{Physical interpretations}


Because similar flow measurement techniques have been applied at Bevalac and RHIC energies with similar motivations it is commonly assumed  that azimuth multipoles have a common source over a broad collision energy range---hydrodynamic flows, collective response to early pressure. The hydro mechanism was proposed as the common element in~\cite{ollitrault} and persists as the lone interpretation of azimuth multipoles in HI collisions to date.

At Bevalac and AGS energies it is indeed likely that azimuth multipoles result from `flow' of initial-state nucleons in response to early pressure, with consequent final-state correlations of those nucleons---a true hydro phenomenon. However, at SPS and RHIC energies the source of azimuth multipoles inferred from final-state produced hadrons (mainly pions) may not be hydrodynamic, in contrast to arguments by analogy with lower energies. Other sources of multipole structure should be considered
\cite{minijetflow1,minijetflow2}. Multipoles at higher energies could arise at the partonic or hadronic level, early or late in the collision, with collective motion or not, and if collective then implying thermalization or not. 


The chain of argument most often associated with elliptic flow asserts that observation of flow as a collective phenomenon demonstrates that a thermalized medium (QGP) has been formed which responds hydrodynamically to early pressure and converts an initial configuration-space eccentricity to a corresponding quadrupole moment in momentum space.

However, nonflow in the form of minijets provides contradictory evidence.  Minijet centrality trends indicate that thermalization is incomplete, and {\em substantial} manifestations of initial-state parton scattering remain at kinetic decoupling~\cite{axialci,ptscale,edep}. Precision studies of minijet centrality dependence ($\nu$ dependence) indicate that a large fraction of the minijet structure expected from {\em linear superposition} of N-N collisions (no thermalization) persists in central Au-Au collisions. That contradiction requires more complete experimental characterization and careful theoretical study~\cite{incompletet}.


Arguments based on interpreting the quadrupole component as hydrodynamic flow exclude alternative physical mechanisms.  Aside from minijet systematics there are other hints that a different mechanism might be responsible for azimuth multipoles. In Fig.~\ref{fig9} we showed that flow measurements based on four-particle cumulants (with bias sources and nonflow thereby reduced) are best described by a trend (solid curves) that is inconsistent with the hydro expectation $v_2 \propto \epsilon$. The trend is instead simply described in terms of per-particle measure $\Delta \rho / \sqrt{\rho_{ref}}$ and two shape parameters relative to $ \epsilon$.

We question the theoretical assumption that $\epsilon$ should be simply related to $v_2$ as opposed to some other measure of the azimuth quadrupole component. We expect {\em a priori} and find experimentally that variance measures, integrals over two-particle momentum space, more typically scale linearly with geometry parameters. Thus, $\Delta \rho[2] / \sqrt{\rho_{ref}} \propto \bar n v_2^2$ may be more closely related to $\epsilon$, and the relation may or may not be characteristic of a hydro scenario.
                     

 \section{Summary}

 In conclusion, we have reviewed Fourier transform theory, especially the relation of autocorrelations to power spectra, essential for analysis of angular correlations in nuclear collisions. In that  context we have reviewed five papers representative of conventional flow analysis and have related the methods and results to autocorrelation structure and spherical and cylindrical multipole moments.

We have examined the need for event-plane evaluation in correlation measurements and find that it is extraneous to measurement of azimuth multipole moments. The EP estimate drops out of the final ensemble average.

We have introduced the definition of the 2D (joint) angular autocorrelation and considered the distinction between flows (cylindrical multipoles) and nonflow (dominated by minijet structure) in conventional flow analysis and criticized the basic assumptions used to distinguish the two in that context.

Based on measured minijet and flow centrality trends we have constructed a simulation exercise in which model autocorrelations of known composition are combined with statistical noise from a typical event ensemble and fit with a model function consisting of a few simple components, first as a 2D autocorrelation and second as a 1D projection on azimuth difference axis $\phi_\Delta$. We show that the 2D fit returns input parameters accurately at the percent level, whereas the 1D fit, representing conventional flow analysis, deviates systematically and strongly from the input. Comparisons with published flow data indicate that the observed bias in the simulation is exactly the difference attributed to ``nonflow'' in conventional measurements.

By comparing our simple algebraic model of quadrupole centrality dependence to data we observe that the trend $v_2 \propto \epsilon$ is not met for any collision system, nor is there asymptotic approach to such a trend. That observation raises questions about the relevance of hydrodynamics to phenomena currently attributed to elliptic flow at the SPS and RHIC.

This work was supported in part by the Office of Science of the U.S. DoE under grant DE-FG03-97ER41020.


\appendix


\section{Brownian motion} \label{brown}

There is a close analogy between Brownian motion and the azimuth structure of nuclear collisions. The long history of Brownian motion and its mathematical description can thus provide critical guidance for the analysis of particle distributions. Brownian motion (more generally, random motion of particles suspended in a fluid) was modeled by Einstein as a diffusion process (random walk)~\cite{brownian}. He sought to test the ``kinetic-molecular'' theory of thermodynamics and provide direct observation of molecules. Paul Langevin developed a differential equation to describe such motion, which included a {\em stochastic} term representing random impulses delivered to the suspended particle by molecular collisions. Jean Perrin and collaborators performed extensive measurements which confirmed Einstein's predictions and provided definitive evidence for the reality of molecules~\cite{perrin}.

\subsection{The quasi-random walker}

We model a 2D quasi-random walker (including nonzero correlations) as follows. The walker position is recorded in equal time intervals $\delta t$. After $n$ steps, with step-wise displacements $r$ sampled randomly from a bounded distribution, the walker position relative to an arbitrary starting point is, in the notation of this paper, $\vec R = \sum_i^n r_i \vec u(\phi_i)$, where $r_i$ is the $i^{th}$ displacement. The squared total displacement is then 
\bea \label{ranwalk}
R^2 = n \langle r^2  \rangle + n(n-1) \langle r^2 \cos(\phi_\Delta) \rangle.
\eea 
The first term, linear in $n$ (or $t$), was described by Einstein. The second term could represent ``drift'' of the walker due to deterministic response to an external influence. The composite is then termed ``Brownian motion with drift,'' a popular model for stock markets and other quasi-random processes. Measuring multipole moments on azimuth in nuclear collisions is formally equivalent to measuring ``drift'' terms on time in the quasi-random walk of a charged particle suspended in a molecular fluid within a superposition of oscillating electric fields. There are many other applications for Eq.~(\ref{ranwalk}).

For a true random walk consisting of uncorrelated steps Einstein expressed $\langle r^2 \rangle / \delta t \equiv d \cdot 2D$ (random walk in $d$ dimensions) in terms of diffusion coefficient $D$. The second term $\langle r^2 \cos(\phi_\Delta) \rangle \equiv (\delta t)^2 v_x^2$ represents a possible deterministic component (correlations), with $\hat x$ the direction of an applied ``force.'' In that case successive angles $\phi_i$ are correlated, and the result is a macroscopic nonstochastic drift of the walker trajectory.

The fractal dimension of a random walk [first term in Eq.~(\ref{ranwalk})] is $d_f = 2$. The trajectory is therefore a ``space-filling'' curve in 2D configuration space. The appropriate measure of trajectory size is area, and the rate of size increase is the diffusion coefficient (rate of area increase). In contrast, the second term in Eq.~(\ref{ranwalk}) represents a deterministic trajectory whose nominal dimension is 1 (modulo the extent of curvature, which increases the dimension above 1). Therefore, the appropriate measure of trajectory size is length, and speed is the correct rate measure. For Brownian motion with drift the trajectory dimension is not well-defined, depending on the relative magnitudes of the drift and stochastic terms, and the concept of speed is therefore ambiguous. Attempts to measure the linear speed of Brownian motion in the nineteenth century failed because of the fractal structure of random walks. From the structure of Eq.~(\ref{ranwalk}) the average speed over interval $\Delta t = n \delta t$ is $\sqrt{R^2 / (\Delta t)^2} \sim   \langle r^2/(\delta t )^2 \rangle / n  $, and the limiting case for $\Delta t = n\delta t \rightarrow 0$ is the so-called ``infinite speed of diffusion.''  That topological oddity is formally equivalent to the ``multiplicity bias'' of conventional flow analysis.

\subsection{Brownian motion and nuclear collisions}

We now consider the close analogy between Einstein's theory of Brownian motion and the measurement of $\overline{p^2_x}$ in a nuclear collision, using directivity as an example. Just as $\vec R$ is the vector total displacement of a quasi-random walker in 2D configuration space, $\vec Q_1$ is the vector total displacement of a quasi-random walker (event-wise particle ensemble) in 2D momentum space. After $n$ steps the squared displacements are
\bea
R^2 = n^2\, \delta t^2\, v'^2_x &=&  n\, \delta t \, 4 D  +  n(n-1) \, \delta t^2 \,v_x^2 \\ \nonumber
Q_1^2 = n^2\, p'^2_x &=&   n\, \langle  p_t^2 \rangle + n(n-1) p_x^2.
\eea
$4 D\delta t$ is the increase in area per step of a random walker in 2D configuration space. $\langle p_t^2 \rangle$ is the increase in area per step (per particle) of a random walker in 2D momentum space, playing the same role as the diffusion coefficient. The RHS first term in the first line is the subject of Einstein's 1905 Brownian motion paper. Its measurement by Perrin confirmed the reality of molecules and the validity of Boltzmann's kinetic theory.  

As noted, attempts to measure mean speed $v'_x$ of a particle in a fluid failed because speed is the wrong rate measure for trajectory size increase. Speed measurements decreased with increasing sample number or observation time. It was not until Einstein's formulation and later mathematical developments that the topology of the random walk and its consequences became apparent. Initial attempts at the Bevalac to measure $p_x$ in the form $p'_x$ using directivity failed for the same reason. Corrections were developed to approximate the unbiased quantity $p_x$, and the failure was attributed to multiplicity bias or `autocorrelations.'  Ironically, the autocorrelation distribution is the ideal method to access the unbiased quantity in either case.

\subsection{Einstein and autocorrelations}

To provide a statistical description of Brownian motion Einstein introduced the autocorrelation concept with the following language~\cite{brownian}.

\begin{quote} 
Another important consideration can be related to this method of development. We have assumed that the single particles are all referred to the same co-ordinate system. But this is unnecessary, since the movements of the single particles are mutually independent. We will now refer the motion of each particle to a co-ordinate system whose origin coincides at the [arbitrary] time $t = 0$ with the [arbitrary] position of the center of gravity of the particle in question; with this difference, that [probability distribution] $f(x,t) dx$ now gives the number of the particles whose $x$ co-ordinate has increased between the time $t = 0$ and the time $t = t$, by a quantity which lies between $x$ and $x + dx$.
\end{quote}

Einstein's function $f(\xi,\tau)$ is a 2D autocorrelation which satisfies the diffusion equation. The solution is a gaussian on $x$ relative to an {\em arbitrary} starting point (thus defining difference variables $\xi = x - x_{start}$ and $\tau = t - t_{start}$), with 1D variance $\sigma^2_\xi = 2 D \tau$. The autocorrelation is sometimes called a {\em two-point correlation function} or  {\em two-point autocorrelation}. The angular autocorrelation is a wide-spread and important analysis tool, e.g., in astrophysics, nuclear collisions and many other fields.

\subsection{Wiener, Khintchine, L\'evy and Kolmogorov}

The names Wiener, L\'evy, Kolmogorov and Khintchine figure prominently in the copious mathematics derived from the Brownian motion problem. Norbert Wiener led efforts to provide a mathematical description of Brownian motion, abstracted to a {\em Wiener process}, a special case of a {\em L\'evy process} (generalization of a discrete random walk to a continuous random process)~\cite{levy}. The Wiener-Khintchine theorem provides a power-spectrum representation for stationary stochastic processes such as random walks, for which a Fourier transform does not exist. We have acknowledged the theorem with our Eq.~(\ref{wienkh}).

The analysis of azimuth structure in nuclear collisions in terms of angular autocorrelations is based on powerful mathematics developed throughout the past century. Autocorrelations make it possible to study azimuth structure for any event multiplicity down to p-p collisions with as little as two detected particles per event. The effects of ``non-flow'' can be eliminated from ``flow'' measurements (and {\em vice versa}) without model dependence or guesswork. The Brownian motion problem and Einstein's fertile solution inform two central issues for studies of the correlation structure of nuclear collisions: analysis methodology and physics interpretation.

\section{Random variables} ~\label{stats}

A random variable represents a set of samples from a {\em parent distribution}. The outcome of any one sample is unpredictable (i.e., random), but through statistical analysis an ensemble of samples can be used to infer properties (statistics -- results of algorithms applied to a set of samples) of the parent distribution. Sums over particles and particle pairs of kinematic quantities are the primary random variables in analysis of nuclear collision data.

\subsection{The algebra of random variables} 

Products and ratios of random variables behave non-intuitively because random variables don't obey the algebra of ordinary variables. E.g., factorization of random variables results in the spawning of covariances.  The approximation $\overline{xy} \simeq \bar x \, \bar y$ common in conventional flow analysis is a source of systematic error (bias) because $\overline{xy} = \bar x \, \bar y + \overline{xy - \bar x\, \bar y}$. The omitted term is a covariance. Such covariances play a role in statistics similar to QM commutators, with $1/n \leftrightarrow \hbar$.  Conventional flow analysis assumes the $1/n \rightarrow 0$ limit for some random variables, and the results are undependable for small multiplicities. Similarly, improper treatment of ratios of random variables results in infinite series of covariances. E.g., 
\bea
\overline{x/n} = \frac{\bar x }{\bar n}\, (1 + \frac{\overline{\delta x\cdot \delta n}}{\bar x\, \bar n} +  + \frac{\overline{x\cdot (\delta n)^2}}{\bar x\, \bar n^2} + \cdots),
\eea 
with $\overline{(\delta n)^2} / \bar n \equiv \sigma^2_n / \bar n \sim 1 - 2$. Thus, the common approximation $\overline{x/n} \simeq \bar x / \bar n$ can result in significant $n$- and physics-dependent ($x$-$n$ covariances) bias for small $n$. 

In this paper we distinguish between event-wise and ensemble-averaged quantities and do not employ ensemble averages of ratios of random variables. We include {\em event-wise} factorizations and ratios only to suggest qualitative connections with conventional  flow analysis. E.g., we consider $\tilde V_m^2 \equiv n(n-1)\langle \cos(m \phi_\Delta) \rangle$ with $\langle \cos(m \phi_\Delta) \rangle = \langle \cos^2(m [\phi - \Psi_r ]) \rangle \equiv \tilde v_m^2$. But, $\overline{\tilde v_m^2} \neq V_m^2 / \overline{n(n-1)} \neq \bar v_m^2$. $v_m$ as typically invoked in conventional flow analysis is not a well-defined statistic. 

\subsection{Statistical references}

The concept of a statistical reference is largely absent from conventional flow analysis. By `statistical reference' we mean a quantity or distribution which represents an uncorrelated system, a system consistent with independent samples from a fixed parent distribution (central limit conditions~\cite{cltpaper}). Concerns about `bias' from low multiplicities~\cite{danod,volzhang,poskvol} typically relate to the presence of an unsubtracted and unacknowledged statistical reference in the final result. Finite multiplicity fluctuations are then said to produce systematic errors, false azimuthal anisotropies, a problem masking true collective effects.

In the limit $1/n \rightarrow 0$ the statistical reference may indeed become negligible compared to the true correlation structure. However, its presence for nonzero $1/n$ is a potential source of systematic error which may block access to important small-multiplicity systems (peripheral collisions and/or small kinematic bins). In general, if the statistical reference is not correctly subtracted the result is increasingly biased with smaller multiplicities. Identification and subtraction of the proper reference is one of the most important tasks in statistical analysis.

Use of the term `statistical' to mean `uncorrelated' is misleading (e.g., `statistical' {\em vs} `dynamical'). All random variables and their fluctuations about the mean are `statistical.' Some random variables and their statistics are {\em reference quantities}, representing systems that are by construction uncorrelated (independent sampling from a fixed parent). We therefore label statistical {reference} quantities `ref,' not `stat.'

\subsection{Random variables and Fourier analysis}

In the context of Fourier analysis the basic finite-number (Poisson) statistical reference is manifested as the delta-function component in the autocorrelation density Eq.~(\ref{autocorr}) and the white-noise constant term $n \langle r^2 \rangle$ in the event-wise power spectrum. Other reference components may arise from two-particle correlations which are not of interest to the analysis (e.g., detector effects) and which may be revealed in mixed-pair distributions. A clear distinction should always be maintained between the reference and the sought-after correlation signal.

Careful attention to random-variable algebra is especially important in a Fourier analysis. The power-spectrum elements and autocorrelation density  must satisfy the transform equations both for each event and after ensemble averaging. In conventional flow analysis that condition is often not satisfied. For instance, $\tilde V_m^2$ and $V_m^2$ satisfy the FT transforms and Wiener-Khintchine theorem before and after ensemble averaging respectively, whereas the ${v_m}$ do not. 

\subsection{Minimally-biased random variables}

It is frequently stated in the conventional flow literature that flow analysis must insure sufficiently large multiplicities. The operating assumption in the design of conventional flow methods is the continuum limit $1/n \rightarrow 0$, with inevitable bias for smaller multiplicities. However, careful reference design and algebraic manipulation of random variables makes possible precise treatment of event-wise multiplicities down to $n = 1$. Some statistical measures perform consistently no matter what the sample number. The full multiplicity range is essential to measure azimuth multipole evolution with centrality down to N-N and p-p collisions, so that A-A ``flow'' phenomena may be connected to phenomena observed in elementary collisions and understood in a QCD context.

Since multiplicity necessarily varies strongly with centrality, multiplicity-dependent bias in flow measurements is unacceptable, and every means should be used to insure minimally-biased statistics. To achieve that end analysis methods must carefully transition from safe event-wise factorizations (as featured in this paper) to ensemble averages minimally biased for {all} $n$. Linear combinations of powers of random variables, e.g., variances and covariances, satisfy a linear algebra. Such integrals of two-particle momentum space are nominally free of bias.

\section{Multipoles and sphericity}

The 1D Fourier transform on azimuth is part of a larger representation of angular structure. The encompassing context is a 2D multipole decomposition on $(\theta,\phi)$ represented by the {\em sphericity} tensor, with the {\em spherical harmonics} $Y_2^m$ as elements. In limiting cases submatrices of the sphericity tensor reduce to ``cylindrical harmonics'' $\cos(m\phi)$, part of the 1D Fourier representation on azimuth.

The central premise of a multipole representation is that the final-state particle angular distribution on $[\theta(y_z),\phi]$ is {efficiently} represented by a few low-order spherical harmonics (SH)  $Y_l^m(\theta,\phi)$. At the Bevalac, sphericity tensor ${\cal S}$ containing  spherical harmonics $Y_2^m$ as elements was introduced. Directivity $\vec Q_1$, simply related to  $Y_2^1$, was employed to represent a rotated quadrupole as a  dipole pair antisymmetric about the collision midpoint. At lower energies (Bevalac, AGS) the quadrupole principal axis may be rotated to a large angle with respect to the collision axis and  $Y_2^1$ dominates.  At higher energies and near midrapidity ($\theta \sim \pi / 2$) the dominant SH is $Y_2^2$. 

\subsection{Spherical harmonics}

The spherical harmonics are defined as
\bea
Y_{lm}(\Omega) = \sqrt{\frac{2l+1}{4\pi}\cdot \frac{(l - m)!}{(l + m)!}}\, P_l^m(\cos \theta)\, e^{i\, m \phi},
\eea
where $P_l^m(\theta)$ is an associated Legendre function~\cite{jackson}.
An event-wise density on the unit sphere can be expanded as
\bea \label{sphereharm}
\tilde \rho(\Omega) &=& \sum_{lm} \tilde {\bf Q}_{lm} \, Y_{lm}(\Omega) \\ \nonumber
\tilde  {\bf Q}_{lm} &=& \int d\Omega \, Y^*_{lm}(\Omega) \tilde \rho(\Omega)  \\ \nonumber
 &=& \sum_i^n Y^*_{lm}(\Omega_i)  \\ \nonumber
&=& n \langle Y^*_{lm}(\Omega) \rangle,
\eea
where $\Omega \rightarrow (\theta,\phi)$ and $d\Omega \equiv d\cos(\theta) d\phi$. The FTs on $\phi$ form a special case of those relations when $\tilde \rho(\Omega)$ is peaked near $\theta \sim \pi/2$. The $Y_{lm}$ are orthonormal and complete:
\bea
\int d\Omega \, Y_{lm}(\theta,\phi)Y_{l'm'}(\theta,\phi) &=& \delta_{ll'}\delta_{mm'} \\ \nonumber
\sum_{l,m} Y_{lm}(\theta,\phi)Y^*_{lm}(\theta',\phi') &=& \delta(\Omega - \Omega').
\eea

\subsection{Multipoles} \label{multi}

The spherical harmonics are model functions for single-particle densities on $(\theta ,\phi)$. The coefficients of the multipole expansion of a distribution are complex {\em spherical} multipole moments describing $2^l$ poles and defined as ensemble averages of the spherical harmonics over the unit sphere weighted by an angular density.

The following relation is defined by analogy with the expansion of an electric potential in spherical harmonics, in this case on momentum space $\vec p$ rather than configuration space $\vec r$~\cite{jackson}
\bea
\int d^3p'\, \frac{\tilde \rho(p',\Omega')}{|\vec p - {\vec p}\, '|} &=& \sum_{lm} \frac{4 \pi}{2l+1} \tilde {\bf q}_{lm} \frac{Y_{lm}(\Omega)}{p^{l+1}}.
\eea
The coefficients are the event-wise {\em spherical multipole moments}
\bea
\tilde {\bf q}_{lm} &\equiv& \int p^2dp\, d\Omega\, p^l  \tilde \rho(p,\Omega) Y^*_{lm}(\Omega) \\ \nonumber
 &=& \sum_i^n p_i^l\,  Y^*_{lm}(\Omega_i) \\ \nonumber
&=& n \langle p^l\,  Y^*_{lm}(\Omega) \rangle.
\eea
Eq.~(\ref{sphereharm}) is the special case for $p$ restricted to unity (i.e., distribution on the unit sphere).

In general, $\Re Y_m^m \propto \sin^m(\theta)\, \cos(m \phi)$, and the $\cos(m \phi)$ are by analogy ``cylindrical harmonics''~\footnote{The term ``cylindrical harmonic'' is conventionally applied to Bessel functions of the first kind, assuming cylindrical symmetry. However, the term ``spherical harmonic'' does not assume spherical symmetry, and we use the term ``cylindrical harmonic'' in that sense by analogy to denote sinusoids on azimuth.}. The ensemble average of a cylindrical harmonic over the unit circle weighted by 1D density $\rho(\phi)$ results in complex {\em cylindrical} multipole moments $ {\bf Q}_m$. The Fourier coefficients  $Q_m$ obtained from analysis of SPS and RHIC data are therefore cylindrical multipole moments describing $2m$ poles. E.g., $m = 2$ denotes a quadrupole moment and $m = 4$ denotes an octupole moment,. 

If nonflow contributions (i.e., structure rapidly varying on $\eta$ or $y$) are present, a multipole decomposition of $\rho(\theta,\phi)$ is no longer efficient, and the inferred multipole moments are difficult to interpret physically (e.g., flow inferences {\em per se} are biased). In Sec.~\ref{joint} we describe a more differential method for representing angular structure using two-particle {\em joint angular autocorrelations} on difference axes $(\eta_\Delta,\phi_\Delta)$.  Given a decomposition of $\rho(\theta,\phi)$ based on variations on $\eta_\Delta$ we can distinguish cylindrical multipoles accurately from ``nonflow'' structure ({\em cf.} Sec.~\ref{autostruct}).

\subsection{Sphericity} \label{sphere}

The sphericity tensor has been employed in both jet physics and flow studies.
A normalized 3D sphericity tensor was defined in~\cite{spherpart} to search for initial evidence of jets in $e^+$-$e^-$ collisions. A decade later sphericity was introduced to the search for collective nucleon flow in heavy ion collisions~\cite{spherflow}. The close connection between flow and jets continues at RHIC, where we seek the relation between minijets and ``elliptic flow.''

Event-wise sphericity $\tilde {\cal S}$ is a measure of structure in single-particle density $\rho(\theta,\phi)$ on the unit sphere. We use dyadic notation to reduce index complexity, analogous to vector notation $\tilde {\vec Q}_m = \sum_i r_i \vec u(m\phi_i)$. $\tilde {\cal S}$ (with $r_i \rightarrow p_i$) describes a 3D quadrupole with arbitrary orientation. Given $\vec p =p\,  [\sin(\theta)\cos(\phi),\sin(\theta)\sin(\phi),\cos(\theta)] \equiv p\, \vec u(\theta,\phi)$ we have
\bea
2 \tilde {\cal S} \equiv 2\sum_{i=1}^n \vec p_i \vec p_i &=& 2 \sum_{i=1}^n p^2_i\, \vec u(\theta_i,\phi_i)\, \vec u(\theta_i,\phi_i)  \\ \nonumber
&=&  \sum_i^n{p_i^2} \, \tilde {\cal U}(\theta_i,\phi_i)   \\ \nonumber
&=& n\langle {p^2} \, \tilde {\cal U}(\theta,\phi) \rangle,
\eea
the last being an event-wise average, where
\bea
&&  {\cal  U}(\theta,\phi) = \sin^2 (\theta)\,  {\cal I} +  {\cal Y}(\theta,\phi)   
\eea
and
\bea
&&  {\cal  Y}(\theta,\phi) =    \\ \nonumber
&& \hspace{-.2in}  \left[
{ \begin{array}{cccc}
&\sin^2(\theta) \cos(2\phi) &\sin^2(\theta)\sin(2\phi)&\sin(2\theta)\cos(\phi)\\
&\sin^2(\theta)\sin(2\phi)&- \sin^2(\theta) \cos(2\phi)&\sin(2\theta)\sin{\phi} \\  
&\sin(2\theta)\cos(\phi)&\sin(2\theta)\sin{\phi}&3\cos^2(\theta) - 1\\
\end{array}} 
\right].
\eea
In terms of event-wise quadrupole moments $\tilde {\bf q}_{2m}$ derived from the $Y_{2m}$ 
\bea
 2\tilde {{\cal S}}{} &=&    n \left\langle p^2\,\sin^2(\theta)\right\rangle  {\cal I}  \\ \nonumber
&+&
 \sqrt{\frac{8}{3}} \, \sqrt{\frac{4\pi}{2l+1}}    \left\langle  \left[
\begin{array}{cccc}
 \Re \tilde {\bf q}_{22}&\Im \tilde {\bf q}_{22}&-\Re  \tilde{\bf q}_{21}\\
\Im  \tilde{\bf q}_{22}&-\Re  \tilde{\bf q}_{22}&-\Im   \tilde{\bf q}_{21} \\  
-\Re   \tilde{\bf q}_{21}&-\Im  \tilde{\bf q}_{21}& \sqrt{\frac{2}{3}}  \tilde{\bf q}_{20}\\
\end{array}
\right] \right \rangle,
\eea
 an event-wise estimator of angular structure on the unit sphere, its reference defined by $2 \tilde {\cal S}_{ref} = n \langle p^2 \sin^2(\theta)  \rangle \,  {\cal I}$, and $p^2 \sin^2(\theta) = p_t^2$. The sphericity tensor of~\cite{spherpart} was normalized to $\hat {\cal S} = {\cal S}  /n \langle p^2 \rangle$.

Note that
\bea
\tilde {\cal Q} = 3 \tilde {\cal S} - n \left\langle p^2 \right\rangle {\cal I}
\eea
is the traceless Cartesian quadrupole tensor appearing in the Taylor expansion of the ($\vec p$ equivalent of the) electrostatic potential~\cite{jackson}. We have defined instead
\bea
\tilde {\cal Q'} = 3 \tilde {\cal S} - \frac{3}{2} n \left\langle p_t^2 \right\rangle {\cal I},
\eea
an alternative quadrupole tensor wherein each element is a single spherical quadrupole moment. The difference lies in the diagonal elements: linear combinations of the $\Re \tilde{\bf q}_{2m}$ in the diagonal elements of $\tilde {\cal Q}$ are simplified to single moments in $\tilde {\cal Q}'$. The ensemble mean of both tensors for an uncorrelated (spherically symmetric) system or system with event-wise quadrupole orientations randomly varying is the null tensor (all elements zero).

\section{Subevents} \label{subev}

The ``subevent'' is a notional re-invention of partitioning/binning, the latter having a history of more than a century in mathematics. In conventional flow analysis subevents are groups of particles in an event segregated on the basis of random selection, charge, strangeness, PID or a kinematic variable such as $p_t$, $y$ or $\eta$. The scalar-product method~\cite{2002} is based on a covariance between two single-particle bins (nominally equal halves of an event). The subevent method is thus a restricted reinvention of a common concept in multiparticle correlation analysis: determining covariances among all pairs of single-particle bins at some arbitrary binning scale -- a two-particle correlation function. Diagonal averages of such distributions are the elements of autocorrelations. 

In the language of conventional flow analysis one way to eliminate statistical reference $Q_{ref}^2$ from $Q_m^2$ is to partition events into a pair of disjoint (non-overlapping) {subevents} A, B~\cite{danod}. In that case $\tilde {\vec Q}_{ma} \cdot \tilde {\vec Q}_{mb} = \tilde {\vec V}_{ma} \cdot \tilde {\vec V}_{mb} = n_a n_b \tilde v^2_{mab}$, a covariance. The partition may be asymmetric (unequal particle numbers) and may be as small as a pair of particles. In addition to eliminating the self-pair statistical reference such partitioning is said to reduce nonflow correlation sources, depending on their physical origins and the partition definition~\cite{2002}. We assume for simplicity that there is no nonflow contribution. Subevent pairs can be used to determine the event-plane resolution for subevents A, B and full events.  

First, we consider the symmetric case, defining equivalent subevents A and B with multiplicities $n_A = n_B = n/2$ from an event with $n$ particles. E.g.,  subevent A has azimuth vector $\vec Q_{mA} = \sum_{i \in A}^{n_A} \vec u(m \phi_i)$. The scalar product is a covariance
\bea
\tilde {\vec Q}_{ma} \cdot \tilde {\vec Q}_{mb} &\equiv& \tilde Q_a\,  \tilde Q_b \langle  \cos(m[\Psi_a - \Psi_b]) \rangle \\ \nonumber
&=& \sum_{i \in A, j\in B}^{n_a, n_b} \cos(m[\phi_i - \phi_j]) \\ \nonumber
&\equiv& n_a\, n_b\, \tilde v_{mab}^2 = \tilde V_{mab}^2,
\eea
with e.g. $\tilde Q^2_a = n_a + n_a(n_a - 1)\tilde v_{ma}^2 = n_a + \tilde V_{ma}^2$. Then
\bea
 \cos(m[\Psi_{ma} - \Psi_{mb}])  &=&  \frac{\tilde V^2_{mab}}{\sqrt{n_a + \tilde V^2_{ma}}\sqrt{n_b + \tilde V^2_{mb}}}.
\eea
If subevents A and B are physically equivalent (e.g., a random partition of the total of $n$ particles), then
\bea
\overline{ \cos(m[\Psi_{ma} - \Psi_{mb}]) } &=& r_{ab}\, \frac{  V^2_{ma}}{\bar n_a +   V^2_{ma}} \\ \nonumber
 && \hspace{-.8in}  = \overline{ \cos(m[\Psi_{ma} - \Psi_r])  \,\cos(m[\Psi_{mb} - \Psi_r]) },
\eea
where $r_{ab} =  V^2_{mab} / \sqrt{ V^2_{ma} \, V^2_{mb}}$ is Pearson's normalized covariance between subevents A and B for the $m^{th}$ power-spectrum elements. If A and B are perfectly correlated ($r_{ab} = 1$) then
\bea
\overline{ \cos(m[\Psi_{ma} - \Psi_{mb}]) } &=& \overline{ \cos^2(m[\Psi_{ma} - \Psi_r])  }
\eea 
In general, $V_m^2 / \bar n = (1 + r_{ab})\,V_{ma}^2 / \bar n_a $, which provides the exact  relation between the EP resolution for subevents and for composite events A + B. It is not generally correct that $\overline{\cos(m[\Psi_m - \Psi_r])} = \sqrt{2}\cdot \overline{\cos(m[\Psi_{ma} - \Psi_r])} $. In this case
\bea
\overline{ \cos^2(m[\Psi_{ma} - \Psi_r])  } = \frac{\bar n_a}{\bar n_a - 1} \cdot \frac{V_{ma}^2}{\bar n_a + V_{ma}^2}
\eea
and $V_{ma}^2 = V_{m}^2 / 4$ for perfectly correlated subevents.

Second, we consider the most asymmetric case A = one particle and B = $n-1$ particles.
\bea
\langle \cos(m[\phi_i - \Psi_r])\, \cos(m[\Psi_{mi} - \Psi_r]) \rangle &=&  \\ \nonumber
&& \hspace{-1.5in} \frac{1}{n} \sum_i \frac{(n-1) \tilde v_{mi}^2}{\sqrt{n-1 + (n-1)(n-2) \tilde v_{mi}^2}} \\ \nonumber
&& \hspace{-1.5in} =\left\langle \tilde v_{m} \cdot  \sqrt{\frac{n-1}{n-2}}  \frac{\tilde V'_m}{\tilde Q'_m}\, \right\rangle,
\eea
where $\tilde Q'^2_m = n-1 + \tilde V'^2_m$ describes a subevent with $n-1$ particles.
In general, the EP resolution for a full event of $n$ particles is given by 
\bea
\overline{ \cos^2(m[\Psi_{m} - \Psi_r]) }  \simeq \frac{\overline{n \tilde V_m^2} }{ \overline{(n-1) \tilde Q^2_m}}.
\eea
Measurement of the EP resolution is simply a measurement of the corresponding power-spectrum element, since
\bea
 V_m^2 / \bar  n \simeq \frac{\overline{ \cos^2(m[\Psi_{m} - \Psi_r]) } }{1 - \overline{ \cos^2(m[\Psi_{m} - \Psi_r]) }}.
\eea
In~\cite{poskvol} the approximation
\bea
\langle \cos(m[\Psi_{m} - \Psi_r]) \rangle^2 &\approx& \frac{\pi}{4} V_m^2 / \bar n
\eea
is given for $V_m^2 / \bar n \ll 1$ or $Q_m^2 \sim \bar n$.

Equal subevents, as the largest possible event partition, imply an expectation that only global (large-scale) variables are relevant to collision dynamics (e.g., to describe thermalized events). The possibility of finer structure in momentum space is overlooked, whereas autocorrelation studies with finer binnings and the covariances among those bins discover detailed event structure highly relevant to collision dynamics.

\section{Centrality issues} \label{centro}

Accurate A-A centrality determination and the centrality dependence of azimuth multipoles and related parameters is critical to understanding heavy ion collisions. We must locate $b=0$ accurately in terms of measured quantities to test theory expectations relative to hydrodynamics and thermalization. And we must obtain accurate measurements for peripheral A-A collisions to provide a solid connection to elementary collisions.

\subsection{Centrality measures}

In~\cite{centmeth} is described the {\em power-law} method of centrality determination. Because the minimum-bias distribution on participant-pair number $n_{part}/2$ goes almost exactly as $(n_{part}/2)^{-3/4}$ the distribution on $(n_{part}/2)^{1/4}$ is almost exactly uniform, as is the experimental distribution on $n_{ch}^{1/4}$, dominated by participant scaling. Those simple forms can greatly improve the accuracy of centrality determination, especially for peripheral and central collisions. The cited paper gives simple expressions for $n_{part}/2$, $n_{bin}$ and $\nu$ relative to fraction of total cross section.

In conventional centrality determination the minimum-bias distribution on $n_{ch}$ is divided into several bins representing estimated fractions of the total cross section. The main source of systematic error is uncertainty in the fraction of total cross section which passes triggering and event reconstruction. The total efficiency is typically 95\%, the loss being mainly in the peripheral region, and the most peripheral 10 or 20\% bins therefore have large systematic errors resulting in abandonment. Flow measurements with EP estimation are also excluded from peripheral collisions due to low event multiplicities. 

In contrast, with the power-law method running integrals of the Glauber parameters and $n_{ch}$ can be brought into asymptotic coincidence for peripheral collisions regardless of the uncertainty in the total cross section. Parameter $\nu$ measures the centrality and greatly reduces the cross-section error contribution. Centrality accuracy $< 2$\% on $\nu$ is thereby achievable down to N-N collisions. That capability is essential to determine the correspondence of A-A quadrupole structure in elementary collisions, to test the LDL hypothesis for instance: is there ``collective'' behavior in N-N collisions?

For central collisions the upper half-maximum point on the power-law minimum-bias distribution provides a precise determination of $b = 0$ on $n_{ch}$ and therefore $\nu$. The $b=0$ point is critical for evaluation of correlation measures relative to Glauber parameter $\epsilon$ in the context of hydro expectations for $v_2 / \epsilon$.

\subsection{Geometry parameters and azimuth structure}

\begin{figure}[h]
 \includegraphics[width=1.65in,height=1.65in]{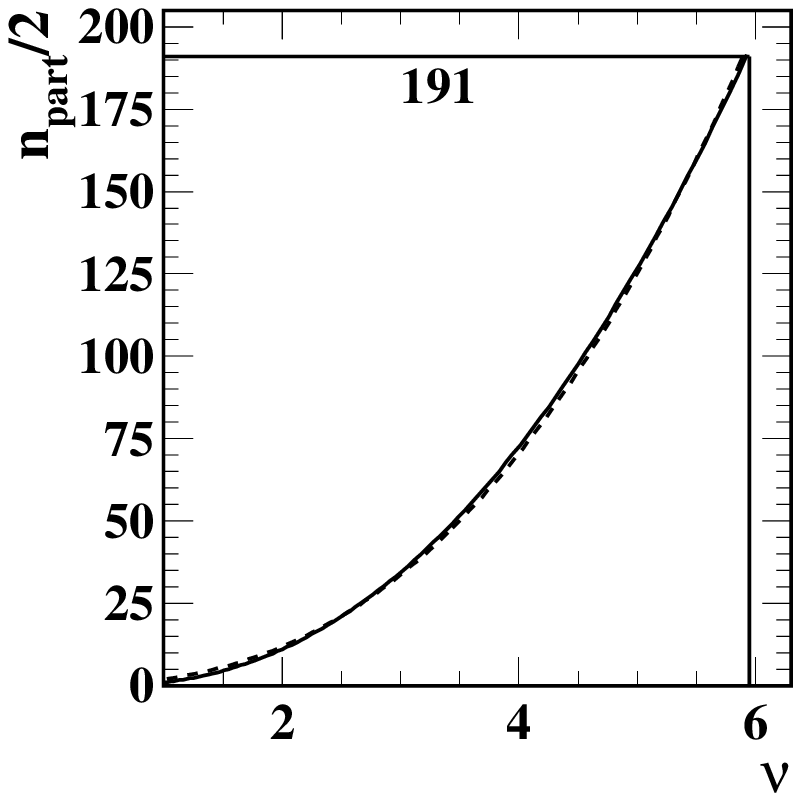}
 \includegraphics[width=1.65in,height=1.65in]{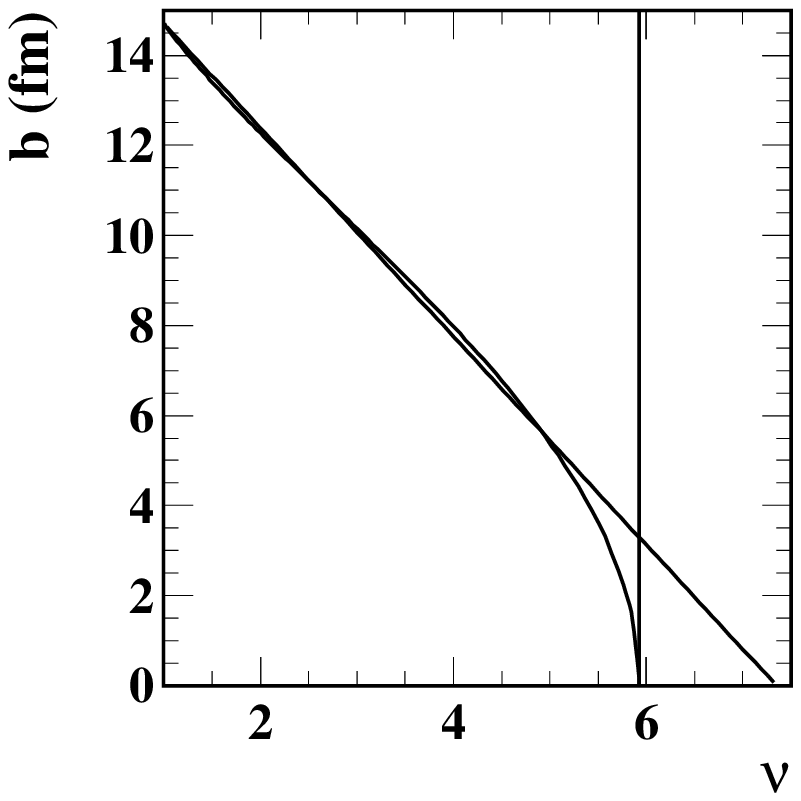}
 \caption{\label{figc1}
Left panel: Participant pair number {\em vs} mean path-length $\nu$ for 200 GeV Au-Au collisions. Because of the nonlinear relation the peripheral third of collisions is compressed to a small interval on $n_{part}/2$. Right panel: Impact parameter $b$  {\em vs} $\nu$. To good approximation the relation is linear over most of the centrality range.
 }  
 \end{figure}

We consider the several A-A geometry parameters relevant to azimuth structure. In Fig.~\ref{figc1} (left panel) we plot $n_{part}/2$ {\em vs} $\nu$ using the parameterization in~\cite{centmeth}. The relation is very nonlinear. The dashed curve is $n_{part}/2 \simeq 2\nu^{2.57}$. The most peripheral quarter of the centrality range is compressed into a small interval on $n_{part}/2$. Mean path-length $\nu$ is the natural geometry measure for sensitive tests of departure from linear N-N superposition, whereas important minijet correlations (nonflow) $\propto \nu$ are severely distorted on $n_{part}$.

In Fig.~\ref{figc1} (right panel) we plot impact parameter $b$ {\em vs} $\nu$, again using the parameterization in~\cite{centmeth} with fractional cross section $\sigma/\sigma_0 = (b/b_0)^2$. We note the interesting fact that over most of the centrality range $b/b_0 \simeq (R - \nu)/(R - 1)$, with $b_0 \equiv 2R = 14.7$ fm for Au-Au. Thus, any anticipated trends on $b$ are also accessible on $\nu$ with minimal distortion.

\begin{figure}[h]
 \includegraphics[width=1.65in,height=1.65in]{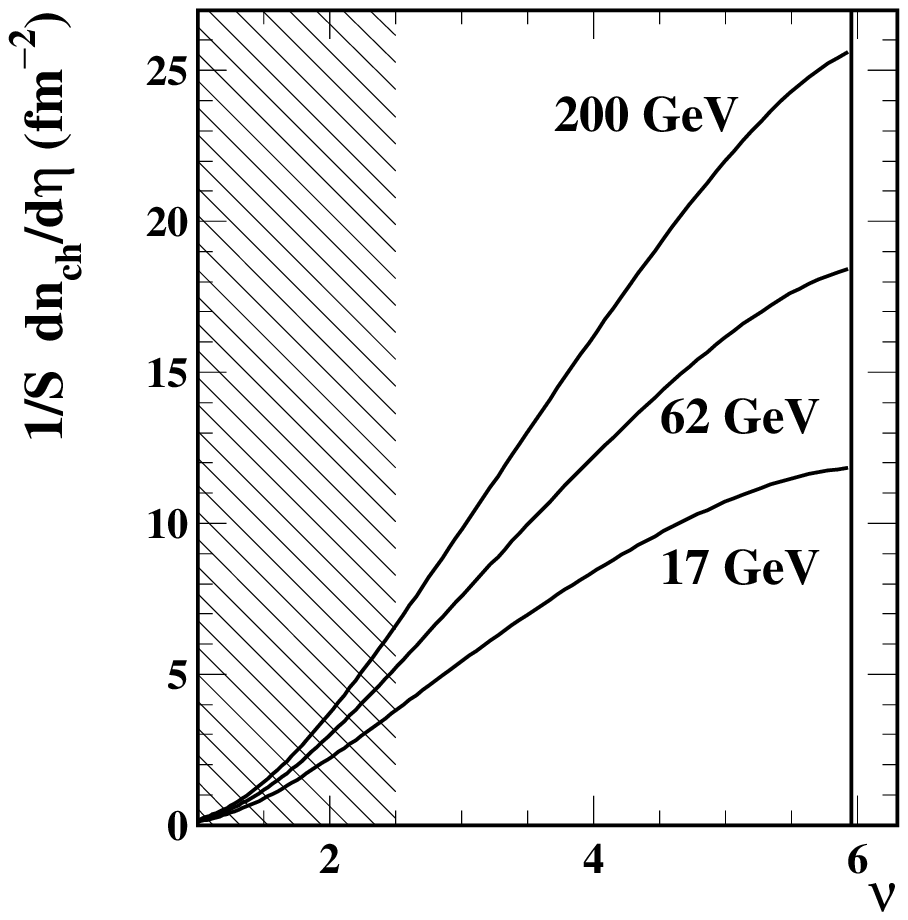}
 \includegraphics[width=1.65in,height=1.65in]{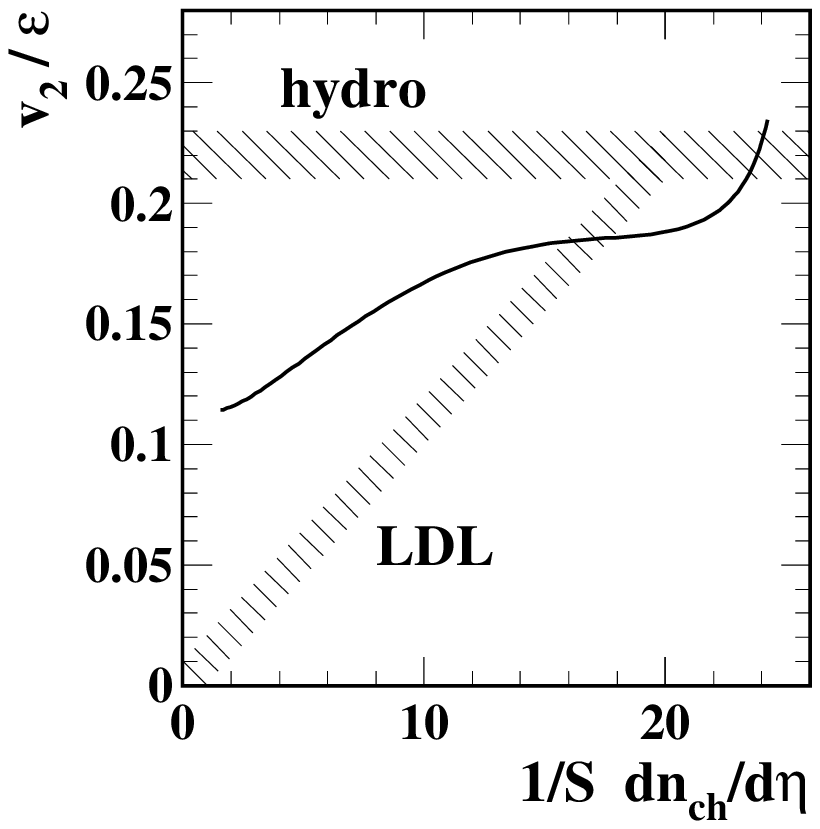}
 \caption{\label{figc2}
 Left panel: Correspondence between LDL parameter $1/S\, dn_{ch}/d\eta$ and centrality measure $\nu$ for three energies. Right panel: Theory expectations for two limiting cases at 200 GeV. The solid curve is derived from the solid curve in Fig.~\ref{fig9} (upper-right panel) using the relation in the left panel.
 }  
 \end{figure}

In Fig.~\ref{figc2} (left panel) we show the LDL parameter $1/S\, dn_{ch}/d\eta$~\cite{heisel} {\em vs} $\nu$ for three collision energies. The energy dependence derives from the multiplicity factor, which we parameterize in terms of a two-component model~\cite{nardi}. Weighted cross-section area $S(b/b_0)$ (fm$^2$) is an optical Glauber parameterization from~\cite{jceps}. Both $\nu$ and $1/S\, dn_{ch}/d\eta$ are pathlength measures. They can be compared with the inverse Knudson number $K_n^{-1}$ introduced in~\cite{incompletet} as a measure of collision number. The LDL measure is based on energy-dependent physical particle collisions, whereas $\nu$ is based on A-A geometry alone. The relation is monotonic and almost linear. Thus, structure on one parameter should appear only slightly distorted on the other.

The hatched region is typically not measured in a conventional flow analysis, due to a combination of large systematic uncertainty in the centrality determination and large biases in flow measurements due to small multiplicities. However, peripheral collisions provide critical tests of flow models: e.g., how does collective behavior (if present) emerge with increasing centrality? In this paper we describe analysis methods which, when combined with the centrality methods of~\cite{centmeth}, make all A-A collisions accessible for accurate measurements down to N-N

In Fig.~\ref{figc2} (right panel) we show $v_2 / \epsilon$ {\em vs} $1/S\, dn_{ch}/d\eta$ for theory expectations (hatched bands) and the simulation in Sec.~\ref{various}. The latter is based on a simple error function on $\nu$ and is roughly consistent with four-particle cumulant results at 200 GeV~\cite{2004}. We observe that the solid curve is not consistent with either the LDL trend for peripheral collisions (the LDL slope is arbitrary) or the hydro trend for central collisions. That provocative result suggests that accurate analysis of azimuth correlations over a broad range of energies and centralities with the methods introduced in this paper and~\cite{centmeth} may produce interesting and unanticipated results.

\subsection{Correlation measures}

If the centrality dependence of azimuth structure is to be accurately determined the correlation measure employed must have little or no multiplicity bias, including statistical biases and irrelevant multiplicity factors which lead to incorrect physical inferences. The quantity $\Delta \rho / \sqrt{\rho_{ref}}$ is the unique solution to a measurement problem subject to multiple constraints. It is the only {\em portable} measure (density ratio) of two-particle correlations applicable to collision systems with arbitrary multiplicity. $\Delta \rho / \sqrt{\rho_{ref}}$ is invariant under linear superposition. If, according to that measure, central Au-Au is different from N-N the difference certainly indicates a unique physical aspect of Au-Au collisions relative to N-N, exactly what we require in a correlation measure. Conventional flow measures do not satisfy that basic requirement. 

Drawing a parallel with measures of $\langle p_t \rangle$ fluctuations we compare $v_2 \leftrightarrow \Sigma_{p_t}$~\cite{ceres}. Both are square roots of per-pair correlation measures  which tend to yield misleading systematic trends (on centrality and energy)~\cite{edep}. In contrast $V_m^2 \leftrightarrow \Delta \Sigma_{p_t:n}^2$~\cite{cltpaper} (the total variance difference for $p_t$ fluctuations) are integrals of two-particle correlations (azimuth number correlations {\em vs} $p_t$ correlations). The first is a measure of total azimuth correlations, the second a measure of total $p_t$ variance difference~\cite{ptfluct}, the integral of a two-particle distribution relative to its reference. In a minimally-biased context we then have $v_m = \sqrt{V_m^2 / \overline{n(n-1)}}$,  analogous to $\Sigma_{p_t}(\text{CERES}) = \sqrt{\Delta \Sigma_{p_t:n}^2 / \overline{n(n-1)}\, \hat p_t^2}$~\cite{edep} (the similarity of notations is coincidental) as the square roots of per-pair measures. Both are physically misleading.
 

\end{document}